\documentclass[draft]{agujournal2019}
\usepackage{url} %this package should fix any errors with URLs in refs.
\usepackage{lineno}
\usepackage[inline]{trackchanges} %for better track changes. finalnew option will compile document with changes incorporated.
\usepackage{siunitx}
\usepackage{soul}
\DeclareSIUnit\ppm{ppm}
\DeclareSIUnit\year{year}
\DeclareSIUnit\years{years}
\draftfalse

\journalname{JGR: Machine Learning and Computation}

\begin{document}

%%%%%%%%%%%%%%%%%%%%%%%%%%%%%%%%%%%%%%%%%%%%%%%
%  TITLE
%
% (A title should be specific, informative, and brief. Use
% abbreviations only if they are defined in the abstract. Titles that
% start with general keywords then specific terms are optimized in
% searches)
%
%%%%%%%%%%%%%%%%%%%%%%%%%%%%%%%%%%%%%%%%%%%%%%%

% Example: \title{This is a test title}
%TC:ignore
\title{Disentangling the effects of sea surface temperature and CO$_2$ in global machine learned weather-climate emulators}

%%%%%%%%%%%%%%%%%%%%%%%%%%%%%%%%%%%%%%%%%%%%%%%
%
%  AUTHORS AND AFFILIATIONS
%
%%%%%%%%%%%%%%%%%%%%%%%%%%%%%%%%%%%%%%%%%%%%%%%

% Authors are individuals who have significantly contributed to the
% research and preparation of the article. Group authors are allowed, if
% each author in the group is separately identified in an appendix.)

% List authors by first name or initial followed by last name and
% separated by commas. Use \affil{} to number affiliations, and
% \thanks{} for author notes.
% Additional author notes should be indicated with \thanks{} (for
% example, for current addresses).

% Example: \authors{A. B. Author\affil{1}\thanks{Current address, Antartica}, B. C. Author\affil{2,3}, and D. E.
% Author\affil{3,4}\thanks{Also funded by Monsanto.}}

\authors{Spencer K. Clark\affil{1,2}, Troy Arcomano\affil{1}, James P. C. Duncan\affil{1}, Brian Henn\affil{1}, Anna Kwa\affil{1}, Jeremy McGibbon\affil{1}, W. Andre Perkins\affil{1}, Elynn Wu\affil{1}, Lucas M. Harris\affil{2}, Oliver Watt-Meyer\affil{1}, and Christopher S. Bretherton\affil{1}}

% \affiliation{1}{First Affiliation}
% \affiliation{2}{Second Affiliation}
% \affiliation{3}{Third Affiliation}
% \affiliation{4}{Fourth Affiliation}

\affiliation{1}{Allen Institute for Artificial Intelligence, Seattle, WA}
\affiliation{2}{NOAA/Geophysical Fluid Dynamics Laboratory, Princeton, NJ}
%(repeat as many times as is necessary)

% Corresponding author mailing address and e-mail address:

% (include name and email addresses of the corresponding author.  More
% than one corresponding author is allowed in this LaTeX file and for
% publication; but only one corresponding author is allowed in our
% editorial system.)

% Example: \correspondingauthor{First and Last Name}{email@address.edu}

\correspondingauthor{Spencer K. Clark}{spencerc@allenai.org}

%%%%%%%%%%%%%%%%%%%%%%%%%%%%%%%%%%%%%%%%%%%%%%%
% KEY POINTS
%%%%%%%%%%%%%%%%%%%%%%%%%%%%%%%%%%%%%%%%%%%%%%%
%  List up to three key points (at least one is required)
%  Key Points summarize the main points and conclusions of the article
%  Each must be 140 characters or fewer with no special characters or punctuation and must be complete sentences

% Example:
% \begin{keypoints}
% \item	List up to three key points (at least one is required)
% \item	Key Points summarize the main points and conclusions of the article
% \item	Each must be 140 characters or fewer with no special characters or punctuation and must be complete sentences
% \end{keypoints}
%TC:endignore
\begin{keypoints}
\item Previous versions of ACE respond unphysically to large independent perturbations of SST or CO$_2$, due to their correlation in training data
\item We mix a subset of previous training data with newly generated data from physics-based simulations with uncorrelated SST and CO$_2$
\item ACE trained on this mixture is more flexible than two previous ACE models combined, while trained on fewer samples than either alone
\end{keypoints}

%%%%%%%%%%%%%%%%%%%%%%%%%%%%%%%%%%%%%%%%%%%%%%%
%
%  ABSTRACT and PLAIN LANGUAGE SUMMARY
%
% A good Abstract will begin with a short description of the problem
% being addressed, briefly describe the new data or analyses, then
% briefly states the main conclusion(s) and how they are supported and
% uncertainties.

% The Plain Language Summary should be written for a broad audience,
% including journalists and the science-interested public, that will not have 
% a background in your field.
%
% A Plain Language Summary is required in GRL, JGR: Planets, JGR: Biogeosciences,
% JGR: Oceans, G-Cubed, Reviews of Geophysics, and JAMES.
% see http://sharingscience.agu.org/creating-plain-language-summary/)
%
%%%%%%%%%%%%%%%%%%%%%%%%%%%%%%%%%%%%%%%%%%%%%%%

%% \begin{abstract} starts the second page

\begin{abstract}

While previous versions of the Ai2 Climate Emulator (ACE) have been trained with CO$_2$ as a forcing, they are only accurate within a narrow range of scenarios, for example climate over the last \num{80} years forced by observed sea surface temperature (SST), sea ice, and CO$_2$ (AMIP), or equilibrium or near-equilibrium climates with CO$_2$ concentrations ranging from 1x to 4x that of the present day.  Attempting to simulate climate forced by AMIP SST perturbed by \SI[retain-explicit-plus]{+4}{\K}, or the response to an abrupt quadrupling of CO$_2$, results in unphysical behavior.  We attribute this to these models being trained on datasets where the SST and CO$_2$ are correlated, limiting their ability to accurately learn their separate effects.  In this study we introduce a new class of ``random-CO$_2$'' reference simulations where the SST and CO$_2$ are prescribed to vary independently.  Trained on a balance of AMIP, equilibrium-climate, and random-CO$_2$ data, and including a total energy conservation constraint for improved interpretability, we present a more data-efficient model that not only accurately emulates its reference model in scenarios in which previous models excelled, but also scenarios like AMIP \SI[retain-explicit-plus]{+4}{\K} and slab-ocean-coupled abrupt 4xCO$_2$ where they did not.  Limitations are that it has simplified or prescribed representations of other Earth system components like the ocean, land, and sea ice; does not expose other known climate drivers as forcings; and relies solely on physics-based model output for training data, inheriting the biases relative to observations thereof.  Each of these represent opportunities for future work.

\end{abstract}

%TC:ignore
\section*{Plain Language Summary}

Machine learned climate models are on the order 100x faster than their physics-based counterparts.  It would be great to be able to use them for climate change experiments, but their flexibility is limited by their training data.  Existing versions of the Ai2 Climate Emulator (ACE) accurately simulate climate when the sea surface temperature and carbon dioxide concentration are roughly in equilibrium, but fail when they are inconsistent with each other.  Classic experiments include uniformly increasing the prescribed sea surface temperature by \SI[retain-explicit-plus]{+4}{\K} without changing the carbon dioxide concentration or abruptly quadrupling the carbon dioxide concentration when running with an interactive ocean.  These experiments help improve physical understanding and demonstrate that models get the right answers for the right reasons.  To address this we train a model using more diverse data, with many different combinations of sea surface temperature and carbon dioxide concentration.  With greater data diversity, we can produce a model that is more flexible than two previous models combined, while trained on fewer samples than either of them alone.  It has comparable accuracy in the scenarios the previous models excelled in, and is also accurate in scenarios where the sea surface temperature and carbon dioxide are varied independently. 
%TC:endignore

% Enter your Plain Language Summary here or delete this section.
% Here are instructions on writing a Plain Language Summary: 
% https://www.agu.org/Share-and-Advocate/Share/Community/Plain-language-summary

%%%%%%%%%%%%%%%%%%%%%%%%%%%%%%%%%%%%%%%%%%%%%%%
%
%  BODY TEXT
%
%%%%%%%%%%%%%%%%%%%%%%%%%%%%%%%%%%%%%%%%%%%%%%%

%%% Suggested section heads:
% \section{Introduction}
%
% The main text should start with an introduction. Except for short
% manuscripts (such as comments and replies), the text should be divided
% into sections, each with its own heading.

% Headings should be sentence fragments and do not begin with a
% lowercase letter or number. Examples of good headings are:

% \section{Materials and Methods}
% Here is text on Materials and Methods.
%
% \subsection{A descriptive heading about methods}
% More about Methods.
%
% \section{Data} (Or section title might be a descriptive heading about data)
%
% \section{Results} (Or section title might be a descriptive heading about the
% results)
%
% \section{Conclusions}

\section{Introduction}

Machine-learning-based weather/climate models provide the ability to generate many realistic realizations of forced weather/climate variability at a fraction of the cost of their traditional physics-based counterparts \cite<e.g.,>{Wat2023, Dun2024, Cha2025a, Cre2025a, Wat2025}.  Models trained on reanalysis data to simulate weather/climate of the recent past, such as ACE2-ERA5 \cite{Wat2025}, have already proven their worth as compelling tools for various scientific questions where generating large ensembles plays an important role \cite<e.g.,>{Chi2026, Lev2026}.

An application where efficient models that coherently simulate a rich representation of the Earth system in space and time could be transformative is in the exploration of different climate change scenarios \cite{Teb2025a}.  Work in this area is still relatively nascent, given the myriad possible forcings, the generalization challenges associated with data-driven models, and coupled nature of the problem, but it represents a natural opportunity for these new tools.  While incorporating observations and/or data from high-resolution physics-based models to arrive at a potentially better estimate of future climate change is a lofty end goal, achieving this in the ``perfect model emulation'' framework, i.e. emulating a single physics-based model, is an important first step, since it would allow for robust validation, and have immediate utility in enabling the exploration of large ensembles of a continuum of scenarios.

\citeA{Cla2025b} provided a start in this direction, by coupling ACE2 to a slab ocean model (SOM), a highly-simplified physics-based ocean used for example in \citeA{Kie2006}, and training it on output from the physics-based SOM-coupled SHiELD model in equilibrium climates with 1x, 2x, and 4x the recent CO$_2$ concentration.  The model, referred to as ACE2-SOM, was able to accurately emulate the characteristics of an out-of-sample intermediate 3xCO$_2$ equilibrium climate, and also reasonably emulate the near-equilibrium climate over the duration of a 70-year simulation with CO$_2$ gradually increasing at a rate of 2\% per year from 1x to 4xCO$_2$.  Where ACE2-SOM fell short, however, was in emulating the response of climate to an abrupt CO$_2$ change.  In an abrupt 4xCO$_2$ test case, with the exception of the thermal-inertia-constrained slab ocean temperature, the state unrealistically shifted nearly immediately to that of a 4xCO$_2$ equilibrium climate in an energy-non-conserving manner.

Another limitation of ACE2-SOM, having been trained only on output from SHiELD coupled to a slab ocean with annually repeating climatological sea ice, is that it is not expected to be able to make accurate predictions when run uncoupled from the SOM and forced with prescribed historical sea surface temperature (SST), sea ice, and CO$_2$.  This SST, particularly earlier in the historical period, is cooler than in any of the climates it was trained on, and also exhibit greater interannual variability; something analogous can be said for sea ice and CO$_2$.  

ACE2-SHiELD, an atmosphere-only model trained on output from prescribed-SST Atmospheric Model Intercomparison Project (AMIP) simulations with SHiELD introduced in \citeA{Wat2025}, accurately emulates historical climate, including during intermediate periods held out from training, but has documented generalization problems of its own, notably predicting land surface cooling in response to uniform SST warming \cite{Wat2025, Zha2026}.  Ideally we would be able to train a model that could accurately emulate both SHiELD and SOM-coupled SHiELD across a range of scenarios with different SST and CO$_2$ forcings.

Since the releases of ACE2-SHiELD and ACE2-SOM, notable work has been done toward better capturing the fast response to an abrupt CO$_2$ change in an ACE-like framework.  \citeA{Mah2026} introduced a hybrid approach where clear-sky vertically-resolved radiative heating perturbations computed using a physics-based radiative transfer scheme were applied to ACE's predictions at each six-hour timestep.  With additional modifications made to the model architecture to increase the locality of its receptive field, and using prescribed SST, they were able to predict the expected responses of precipitation and latent heat flux to abrupt changes in CO$_2$.  This architecture-related insight is powerful, though it has not been demonstrated that this approach would generalize in long simulations if the SST were allowed to respond to the clear-sky radiative heating perturbations, and it does not account for the impact of the fast responses of clouds, like those described in \citeA{Zel2013}.

In this study we introduce a model that can not only accurately perform traditional AMIP and equilibrium-climate SOM-coupled simulations with varying concentrations of CO$_2$ as in previous ACE models, but also accurately perform AMIP simulations with uniform sea surface temperature perturbations and SOM-coupled simulations with abrupt changes in CO$_2$. It does this while being trained on \SI{\sim 25}{\percent} fewer samples than ACE2-SHiELD or ACE2-SOM. We achieve this by training on a combination of reference data from traditional AMIP and SOM-coupled equilibrium-climate runs, in addition to a new class of reference runs with uncorrelated sea surface temperatures and CO$_2$.  For improved sharpness of predictions and representation of extremes, we adopt the stochastic version of ACE introduced in \citeA{Per2026}, and for improved interpretability we include a corrector analogous to that in \citeA{Cha2025a} to ensure that global energy is conserved to the extent that it is in the reference model.

\section{Methods}

\subsection{Reference data}

As reference data for this study, we build upon the catalog of coarse-resolution simulations our group has completed using the SHiELD model developed at the NOAA Geophysical Fluid Dynamics Laboratory (GFDL).  SHiELD is a development version of the United States government's weather forecast model \cite{Zho2019, Har2020}, and has been used in its C3072 (\SI{\sim 3}{\km}) configuration for climate-timescale studies \cite<e.g.,>{Che2022, Har2023, Mer2024}.  The coarse-resolution configuration our group has used as a target for emulation, chosen for computational efficiency reasons, employs a C96 (\SI{\sim 100}{\km}) grid in the horizontal and \num{79} finite-volume hybrid-sigma-pressure levels in the vertical.  With the exception of the ocean boundary conditions, forcing data, and initial conditions used, all of these simulations were run with identical namelist configurations and physics-related code.  For use with and comparison to ACE, all output was regridded to a \SI{1}{\degree} Gaussian grid, spherical harmonic transform roundtrip filtered, and vertically coarsened to ACE's \num{8} finite volume vertical layers, following the procedures described in \citeA{Wat2025}.

These reference simulations can be classified under three categories: AMIP and SOM-coupled, which are carried over largely from \citeA{Wat2025} and \citeA{Cla2025b}, as well as a new ``ramped-SST-random-CO$_2$'' category.  We will describe the simulations in each of these in detail below.  Since we are interested in a model that is capable of both traditional AMIP and SOM-coupled inference with varying CO$_2$, we train new models on a mixture of data from these three categories.  Training data from the AMIP and SOM-coupled categories is drawn from subsets of the same simulations as it was for the previous ACE2-SHiELD and ACE2-SOM models, i.e. a traditional AMIP initial condition ensemble and equilibrium-climate SOM-coupled simulations with 1x, 2x, and 4x the recent concentration of CO$_2$.  Held out simulations which modify the SST and CO$_2$ in various ways serve as out-of-sample tests.  Through ablation experiments, we will show that the inclusion of ramped-SST-random-CO$_2$ data during training is necessary for accurately disentangling the effects of SST and CO$_2$.

\subsubsection{AMIP simulations}

Our AMIP simulation catalog consists of one traditional initial condition ensemble and two perturbation experiments.  For \citeA{Wat2025} we ran a two-member initial-condition ensemble forced by observed SST, sea ice, and carbon dioxide concentration from 1940 through 2020. As a perturbation experiment we also ran a simulation with sea surface temperatures uniformly perturbed by \SI[retain-explicit-plus]{+4}{\K} starting from an initial condition from 1979-01-01 of the second traditional AMIP ensemble member and run through 2020.  For this study, we run an additional perturbation simulation, again starting on 1979-01-01 and running through 2020, where the sea surface temperatures are unperturbed, but the CO$_2$ concentration is held fixed at the 1979 level.  The members of the traditional initial condition ensemble are fully spun up and meteorologically diverged; in the case of both perturbation runs, we treat the first year as spin-up and focus our analysis on the period from 1980 through 2020.

\subsubsection{SOM-coupled simulations}

Our SOM-coupled simulation catalog consists of standalone and ensemble simulations in equilibrium, near-equilibrium, and non-equilibrium climates.  For \citeA{Cla2025b} we ran a five-member initial-condition ensemble of \num{10}-year slab-ocean-model-coupled runs in equilibrium climates with 1x, 2x, 3x, and 4x the year-1997-observed CO$_2$ concentration, \SI{363.43}{\ppm}.  We also ran a \num{70}-year simulation with CO$_2$ increasing at a rate of 2\% per year (2pctCO$_2$), and a \num{10}-year simulation with CO$_2$ abruptly quadrupled.  For this study we add a second \num{70}-year 2pctCO$_2$ ensemble member, and run a \num{36}-member \num{90}-day abrupt 4xCO$_2$ initial condition ensemble.  In the case of the abrupt 4xCO$_2$ ensemble, initial conditions were derived from 00Z on the first day of the initial \num{36} months of the fifth 1xCO$_2$ equilibrium-climate ensemble member.  The equilibrium-climate and 2pctCO$_2$ simulations are all considered spun-up and meteorologically diverged for the durations cited here; we naturally do not ignore any spin up period in the case of the abrupt 4xCO$_2$ simulations.

\subsubsection{Ramped-SST-random-CO$_2$ simulations}

Both ACE2-SHiELD and ACE2-SOM were trained only on data where the SST and CO$_2$ concentration were mutually correlated. Thus it is not surprising that they generalized poorly to situations where the SST and CO$_2$ concentration were not similarly correlated.  In this study we introduce a new class of reference simulations, which we refer to as ``ramped-SST-random-CO$_2$'' runs, or simply ``random-CO$_2$'' runs for short, meant to efficiently fill this gap in the sample space.  This is a \num{9}-member ensemble of \num{5}-year (post spin up) simulations, all forced by the same prescribed SST and sea ice evolution, but with an unique and highly variable CO$_2$ concentration.  

\begin{figure}
\noindent\includegraphics[width=\textwidth]{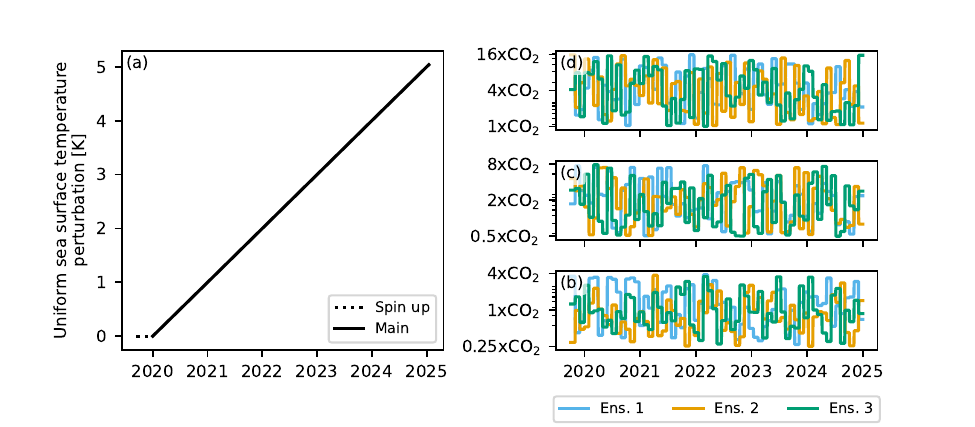}
\caption{Time-varying uniform sea surface temperature perturbation (a) and CO$_2$ concentration logarithmically centered about 1x (b), 2x (c), and 4x (d) the year-1997 CO$_2$ concentration in each ensemble member of the random-CO$_2$ simulations.}
\label{fig:ramped-sst-random-co2}
\end{figure}

The prescribed SST and sea ice are derived from the same annually repeating 1982 to 2012 climatology \cite{Thi2003, Sah2014} used in the run for generating the slab ocean model Q-flux forcing in \citeA{Cla2025b}.  However, for these runs we also superpose a time-varying uniform SST perturbation, illustrated in Figure~\ref{fig:ramped-sst-random-co2}a, which is zero for a three month spin up period to allow for meteorological divergence across the ensemble members, and then increases at a rate of \SI{1}{\K \per \year}.  To sample the matrix of possible combinations of SST and CO$_2$, three ensemble members each were run with the CO$_2$ concentration logarithmically centered about 1x, 2x, and 4xCO$_2$.  To further ensure sampling of non-equilibrium climate states, particularly in the stratosphere, the global CO$_2$ concentration is prescribed following:
\begin{equation}
X(t) = 2^{n(t)} X_o,
\end{equation}
where the exponent $n(t)$ is drawn randomly from a real-valued uniform distribution between \num{-2} and \num{2} every \num{30} days, and $X_o$ is the central CO$_2$ concentration.  This logarithmic centering is chosen due to the observed approximate dependence of radiative forcing on CO$_2$ concentration \cite<e.g.,>{Hua2014a}. The length of time between CO$_2$ changes is chosen primarily to allow for the stratospheric temperature to roughly equilibrate to each new concentration, but may also help in sampling other fast atmospheric adjustments like responses of the hydrological cycle or clouds \cite{Kam2015, Blo2021, Mah2026}.  Figures~\ref{fig:ramped-sst-random-co2}b through ~\ref{fig:ramped-sst-random-co2}d show the time series of the CO$_2$ concentration for each ensemble member, separated by the central CO$_2$ concentration for visual clarity.

\subsection{Machine learning model architecture and training procedure}

In contrast to ACE2-SHiELD and ACE2-SOM, which were trained using the deterministic Spherical Fourier Neural Operator (SFNO) machine learning architecture \cite{Bon2023} with two timesteps in the loss, we train all models for this study using a slightly modified version of ACE2S, which uses a stochastic SFNO architecture and weighted Continuous Ranked Probability Score and Energy Score loss described in \citeA{Per2026}.   We follow a similar training procedure to that of \citeA{Per2026}, beginning with one-step pre-training for \num{30} epochs, followed by multi-step fine-tuning from the one-step pre-training checkpoint with the lowest validation loss for another \num{30} epochs. For multi-step fine-tuning, we use a randomly selected number of steps per batch between \num{1} and \num{20} following the distribution in \citeA{Per2026}, optimizing on only the last predicted step for each sample based on an ensemble size of \num{2}. Using a stochastic model facilitates the generation of large well-calibrated ensembles, and improves the sharpness of predictions, a well-known weakness of deterministic models with a mean-square-error loss \cite{Lan2026b, Bon2025, Per2026}.  Text~S1 and Figure~S1 illustrate the impact of the sharpness improvements with the stochastic version of ACE on the power spectrum and probability distribution function of precipitation in SOM-coupled 3xCO$_2$ inference.

The hyperparameters used to define the model and govern its training dynamics are the same as those used in \citeA{Per2026}, with the exception of increasing the size of the embedding dimension from \num{384} to \num{512}, using a better conditioned weight initialization approach closer to that of \citeA{Bon2025}, and the inclusion of elementwise affine parameters in the normalization layers. The larger embedding dimension size increases the capacity of the model, but comes at the cost of reduced training and inference speed.  When using an embedding dimension of \num{512} versus \num{384} the throughput is roughly \num{12} samples per second versus \num{15} samples per second during multi-step fine-tuning on 8 NVIDIA H100 GPUs and roughly \num{775} simulated years per day versus \num{1150} simulated years per day during inference on a single H100 GPU.  We have not carefully ablated whether it makes an important difference on this specific problem, but when training on the ERA5 reanalysis dataset \cite{Her2020}, using an embedding dimension of \num{512} instead of \num{384} resulted in a meaningful reduction of the error of the time-mean spatial pattern of the predicted variables in five-year rollouts with forcings held out from training.

We use the same variable set as that of \citeA{Wat2025} and \citeA{Cla2025b}, but add to it the total frozen precipitation rate as a diagnostic variable, and switch the CO$_2$ concentration from a same-timestep forcing to a next-timestep forcing.  The former change is to facilitate more accurately computing, and optionally constraining, the global atmosphere total energy budget.  The latter is to address the fact that the CO$_2$ forcing should represent the mean value over the time interval of which we are computing the state update and associated mean boundary energy fluxes, as it does in SHiELD.  This was not particularly important in the ACE2-SHiELD and ACE2-SOM models, where the CO$_2$ varies slowly, if at all, in training and inference rollouts, but in models trained including data from the random-CO$_2$ runs it is relevant at moments where the CO$_2$ abruptly changes.

\subsection{Physical constraints}

In this study we experiment with the inclusion of a global energy conservation constraint, analogous to that introduced in \citeA{Cha2025a}, which can be applied after this base sequence of corrections.  Motivation for including an energy conservation constraint comes in part from the clear energy non-conservation exhibited by ACE2-SOM in abrupt 4xCO$_2$ inference \cite{Cla2025b}, and in general for improved interpretability.  In contrast to \citeA{Cha2025a}, our formulation, described in detail in \ref{sec:global-energy-corrector}, uses a slightly different definition of total energy, and adjusts the temperature at the end of each timestep by adding a global constant rather than with a multiplicative weighting.  All models we train also incorporate the same positive flux and mixing ratio, global dry air conservation, global moisture conservation, and column moisture conservation constraints as those introduced in \citeA{Wat2025}, adding total frozen precipitation rate to the list of variables forced to be positive.

We train models with and without the energy conservation constraint to measure its impact.  When training models with the energy conservation constraint, we only introduce it during the multi-step fine-tuning phase to ensure ACE is already making skillful predictions of the variables that enter the energy budget.  The other physical constraints are applied starting in the one-step pre-training phase.

\subsection{Experimental setup}

Starting from two different random weight initializations, we train models with four different configurations in this study, producing a total of eight models.  Along one dimension we explore the impact of including data from the random-CO$_2$ simulations during training.  Along the other dimension we explore the impact of including the energy conservation constraint.  Table~\ref{tab:data-configurations} in \ref{sec:data-configurations} defines the two data configurations we use when training models, one with just AMIP and equilibrium-climate data, and one with AMIP, equilibrium-climate.  For the remainder of this paper, we consider our main configuration to be the one that includes random-CO$_2$ data in training and the energy conservation constraint, which we label simply as ``ACE2S-SHiELD+.''  For conciseness, we label models trained for ablation experiments with subscripts that denote the absence of particular features, e.g. ``ACE2S-SHiELD+$_{\text{no-RC}}$'' means ACE2S-SHiELD+ trained without random-CO$_2$ data, but including energy conservation; ``ACE2S-SHiELD+$_{\text{no-EC}}$'' means ACE2S-SHiELD+ trained with random-CO$_2$ data, but without energy conservation; and ``ACE2S-SHiELD+$_{\text{no-RC-no-EC}}$'' means ACE2S-SHiELD+ trained without random-CO$_2$ and without energy conservation.

For models trained without random-CO$_2$ data, normalization statistics are derived from the first ensemble members in each of the 1x, 2x, and 4xCO$_2$ equilibrium climates, while for models trained with random-CO$_2$ data, they are derived from the first two ensemble members of the random-CO$_2$ runs across all central CO$_2$ concentrations.  We are careful that the two configurations draw samples from the same total number of years during training, in this case \num{90} years, which is less than the \num{\sim 132} years of AMIP data used to train ACE2-SHiELD \cite{Wat2025} or the \num{120} years of 1x, 2x, and 4xCO$_2$ equilibrium-climate data used to train ACE2-SOM \cite{Cla2025b}.  In the case of the models trained with just AMIP and equilibrium-climate data, the absence of random-CO$_2$ data is filled by equal amounts of AMIP and equilibrium-climate data from additional ensemble members.  Both configurations use the same \num{6} held-out years of AMIP and equilibrium-climate data for validation, while the models trained including random-CO$_2$ data further include an additional \num{3} years of held-out random-CO$_2$ data.  Finally, to aid in selecting a final checkpoint, both configurations run an ensemble of \num{24} \num{5}-year rollouts across the 1x, 2x, and 4xCO$_2$ equilibrium climates after each epoch, which we refer to as ``inline inference.''  The final checkpoints during multi-step fine-tuning are chosen based on the lowest root mean square error of the time and ensemble mean averaged across all normalized output variables, following the same approach described in \citeA{Wat2025}. 

To evaluate generalization, we focus on inference with forcings held out from training.  This includes AMIP from 2012 to 2020, 3xCO$_2$ equilibrium climate, CO$_2$ increasing at a rate of \SI{2}{\percent \per \year}, which are scenarios where the baselines of ACE2-SHiELD and ACE2-SOM do well, and AMIP constant CO$_2$, AMIP \SI[retain-explicit-plus]{+4}{\K}, random-CO$_2$ inference with a held-out CO$_2$ time series, and abrupt 4xCO$_2$, which are scenarios that test how accurately models disentangle the effects of SST and CO$_2$.  For visual clarity, for a given configuration we only show results from the seed which had the lowest inline inference error, but are careful to only draw conclusions that appear robust across random seeds.

Regardless of whether we include random-CO$_2$ data, we train on a mixture of data from prescribed SST and SOM-coupled reference simulations.  However, for simplicity we always train and run inline-inference in prescribed-SST mode.  All relevant SOM-coupled evaluation tests, i.e. equilibrium-climate, 2pctCO$_2$, abrupt 4xCO$_2$, are run zero-shot, i.e. without any fine-tuning, in SOM-coupled mode.  While this differs from what was done in \citeA{Cla2025b}, we find it to be reasonable, since the short rollouts involved during training, at most five days in this case, would offer little time for major deviations to develop in the slab ocean temperature from its initial condition.  What is most important for driving the slab ocean is being able to make accurate predictions of the surface energy fluxes given a particular slab ocean temperature and atmospheric state, which data-ocean training is sufficient to optimize.

\section{Results}

We will begin with a brief discussion of results illustrating that ACE2S-SHiELD+ performs comparably to the two relevant baseline models in their strongest domains: ACE2-SHiELD in traditional AMIP inference, and ACE2-SOM in 3xCO$_2$ equilibrium-climate and 2pctCO$_2$ slab-ocean inference.  We will then focus more in depth on evaluations that illustrate that this model, unlike the baselines, has also learned to accurately disentangle the independent effects of SST and CO$_2$ forcing, with some discussion of the dataset and energy conservation ablation tests.

\subsection{Standard test cases with correlated SST and CO$_2$}

\subsubsection{AMIP inference}

\begin{figure}
\noindent\includegraphics[width=\textwidth]{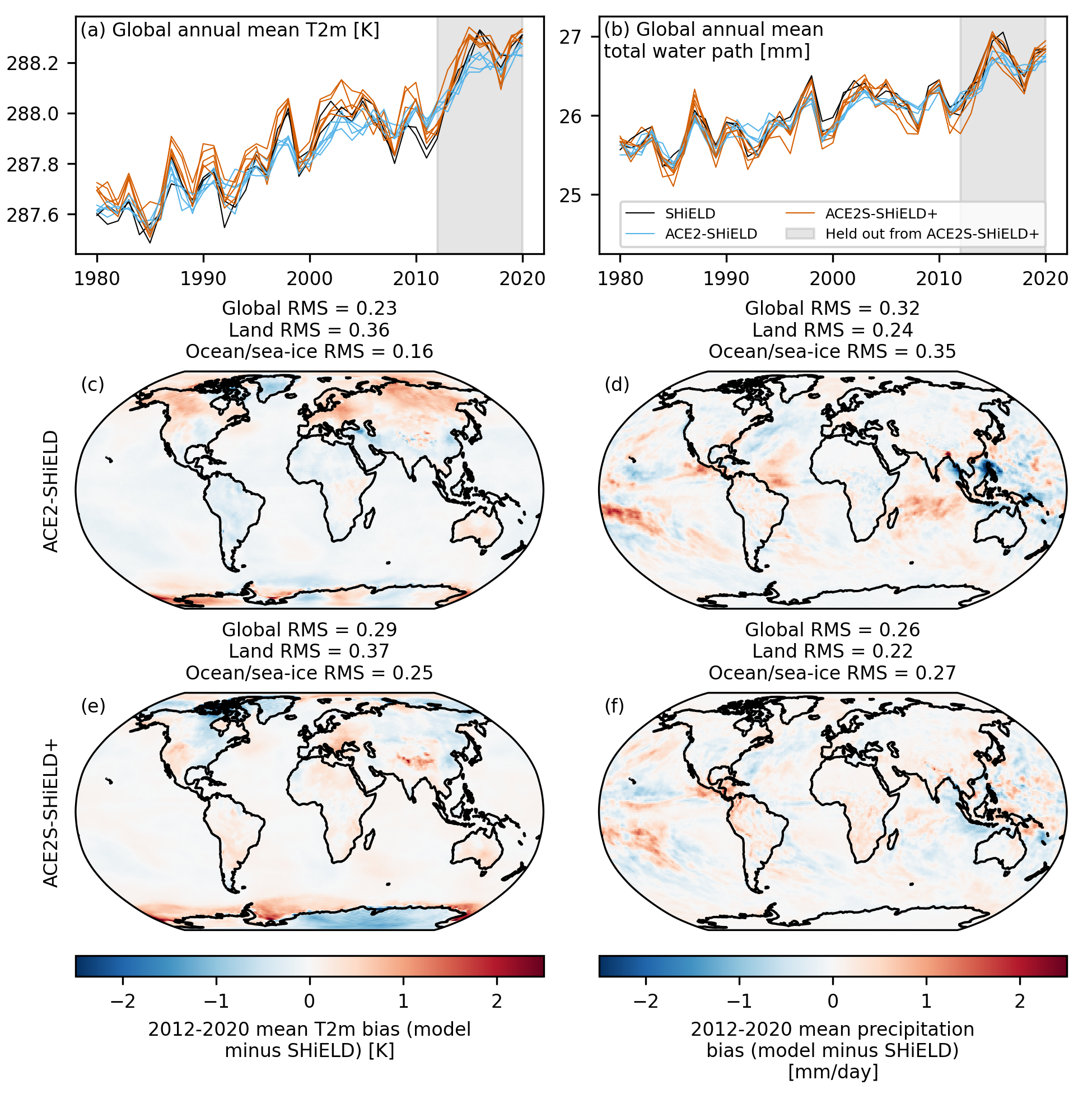}
\caption{Global annual mean time series of \SI{2}{\m} temperature (a) and total water path (b) in an ensemble of simulations with SHiELD, ACE2-SHiELD, and ACE2S-SHiELD+ between 1980 and 2020.  The gray shaded region indicates the region of the dataset held out from training of ACE2S-SHiELD+, though note it was included in training ACE2-SHiELD.  Bias maps of 2012-2020-mean \SI{2}{\m} temperature and precipitation rate for ACE2-SHiELD (c)-(d) and ACE2S-SHiELD+ (e)-(f).  Biases are shown between the first member of each ensemble.}
\label{fig:amip}
\end{figure}

Figure~\ref{fig:amip} compares AMIP inference results with the baseline ACE2-SHiELD model and ACE2S-SHiELD+ trained including random-CO$_2$ data with energy conservation.  Starting at the beginning of 1979, we ran five-member ensembles with both models, discarding the first year as spin up.  The time series plots of global annual mean \SI{2}{\m} temperature (Figure~\ref{fig:amip}a) and total water path (Figure~\ref{fig:amip}b) show data from each individual ensemble member from 1980 through 2020, while the maps show the time-mean bias of \SI{2}{\m} temperature and precipitation rate for the first ensemble member in the period held out from training/validation in ACE2S-SHiELD+, i.e. 2012 through 2020.  Note that this period was included in training ACE2-SHiELD, which only makes this comparison a more challenging test.

ACE2S-SHiELD+ exhibits greater ensemble spread than the deterministic baseline ACE2-SHiELD as measured by the standard deviation of global annual means across the five ensemble members in each year, \SI{40}{\percent} greater for \SI{2}{m} temperature and \SI{51}{\percent} greater for total water path.  It also has comparable $R^2$ when computed over the annual global mean time series from 1980 through 2020 for both \SI{2}{\m} temperature (\num{0.81}-\num{0.93} versus \num{0.86}-\num{0.88}) and total water path (\num{0.82}-\num{0.95} versus \num{0.78}-\num{0.82}).  During the holdout period, ACE2S-SHiELD+ is roughly unbiased in the global mean, which suggests that it has learned from the warmer climates it was exposed to during training even though it had not seen those exact combinations of SST and CO$_2$; models trained only on historical AMIP simulations or reanalysis can exhibit cold biases in this kind of extrapolation problem \cite{Hen2026, Lan2026}.  It does, however, have slightly larger pattern error for emulating the time-mean pattern of \SI{2}{\m} temperature, which appears to be largest in regions of sea ice, like the Antarctic coast and the Canadian Arctic Archipelago.  This may be due to seeing limited sea ice variability during training, since the only samples in which it deviates from a prescribed annually repeating climatology are in the period of the AMIP dataset it was trained on.  

With all common variables considered, i.e. excluding total frozen precipitation rate, ACE2S-SHiELD+ has comparable skill to ACE2-SHiELD in emulating the 2012-2020-mean spatial pattern.  Figure~S2 shows global root mean square error for each predicted variable of the first two ensemble members relative to the two ensemble members from SHiELD.  With the exception of the surface temperature, $T_1$, $T_2$, $q_0$, $q_1$, upward shortwave radiative flux at the surface, upward longwave radiative flux at the surface, and \SI{2}{\m} temperature, the RMSE for ACE2S-SHiELD+ is within the uncertainty range of that for ACE2-SHiELD or lower; here the uncertainty is determined based on the uncertainty of the noise floor, computed following a similar method to that in \citeA{Cla2025b}.

\subsubsection{Equilibrium climate inference}

\begin{figure}
\noindent\includegraphics[width=\textwidth]{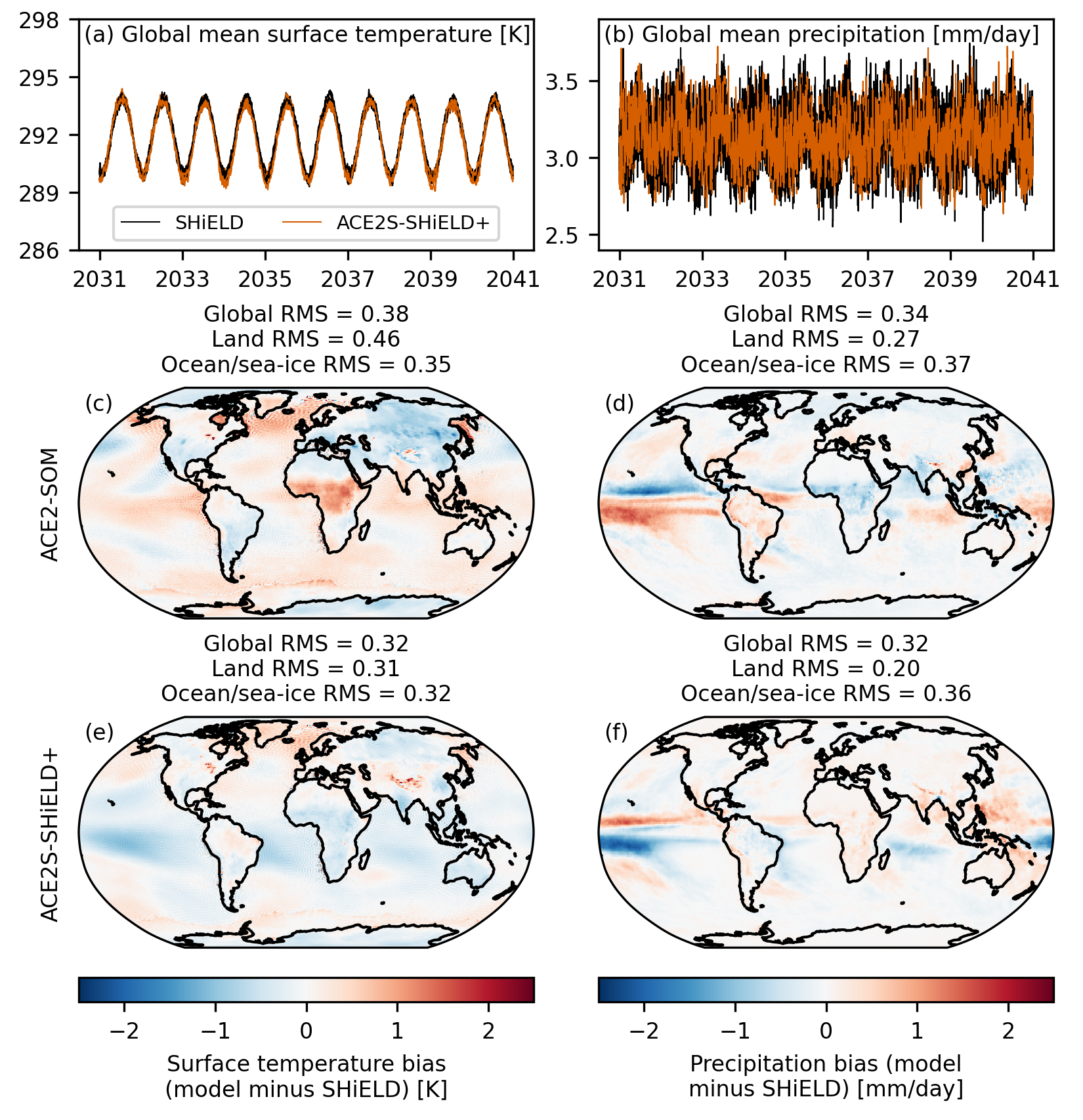}
\caption{Global daily mean time series of surface temperature (a) and precipitation rate (b) in a five-member ensemble of \num{10}-year SOM-coupled simulations with SHiELD (black) and ACE2S-SHiELD+ in a 3xCO$_2$ equilibrium climate.  Bias maps of time and ensemble mean surface temperature and precipitation rate for ACE2-SOM (c)-(d) and ACE2S-SHiELD+ (e)-(f).}
\label{fig:3xCO2-equilibrium-climate}
\end{figure}

Figure~\ref{fig:3xCO2-equilibrium-climate} illustrates the skill of ACE2S-SHiELD+ in SOM-coupled 3xCO$_2$ equilibrium-climate inference, a scenario held out from training.  We ran a five-member ensemble of \num{10}-year simulations with the target SHiELD, the new ACE2S-SHiELD+, and the baseline ACE2-SOM models.  Figures~\ref{fig:3xCO2-equilibrium-climate}a and \ref{fig:3xCO2-equilibrium-climate}b show the time series of global mean surface temperature and precipitation rate for each ensemble member with SHiELD and ACE2S-SHiELD+.  All ensemble members with ACE2S-SHiELD+ are roughly unbiased and exhibit no meaningful drift.  Figures~\ref{fig:3xCO2-equilibrium-climate}c through \ref{fig:3xCO2-equilibrium-climate}f compare the time and ensemble mean biases for surface temperature and precipitation rate between the baseline ACE2-SOM (\ref{fig:3xCO2-equilibrium-climate}c and \ref{fig:3xCO2-equilibrium-climate}d) and ACE2S-SHiELD+ (\ref{fig:3xCO2-equilibrium-climate}e and \ref{fig:3xCO2-equilibrium-climate}f).  In this particular climate and for these variables, ACE2S-SHiELD+ happens to exhibit smaller pattern errors than ACE2-SOM.  When normalized by the noise floor and averaged across all variables and the 1x, 2x, 3x, and 4xCO$_2$ climates, ACE2S-SHiELD+ has comparable skill to ACE2-SOM at emulating the time-and-ensemble-mean spatial pattern of SHiELD.  Figure~S3 illustrates this for the out-of-sample 3xCO$_2$ equilibrium climate, though note that despite it being within the training dataset, this particular checkpoint happens to have worse skill in the 2xCO$_2$ climate.

\citeA{Wat2025} showed that ACE2-ERA5 ran stably for \num{1000} years when forced by annually repeating climatological SST, sea ice, and CO$_2$.  While not shown in \citeA{Cla2025b}, ACE2-SOM is also capable of stable \num{1000}-year SOM-coupled inference in the 1x, 2x, 3x, and 4xCO$_2$ equilibrium climates.  In a similar experiment, ACE2S-SHiELD+ was stable in \num{1000}-year runs in three out of four climates, but exhibited a large regime shift originating in the stratosphere that affected other variables in year \num{506} of the remaining one; more specifically, a fluctuation occurred in the stratospheric northward wind, global mean stratospheric specific total water dropped to near zero within a year, and then global annual mean two meter temperature increased by \SI{1.6}{\K} over the span of about \num{8} years.  More careful work will need to be done to understand the source of this behavior, but ablation tests suggest that the inclusion of random-CO$_2$ data in training or energy conservation are not the principal cause.

\subsubsection{2pctCO$_2$ inference}

\begin{figure}
\noindent\includegraphics[width=\textwidth]{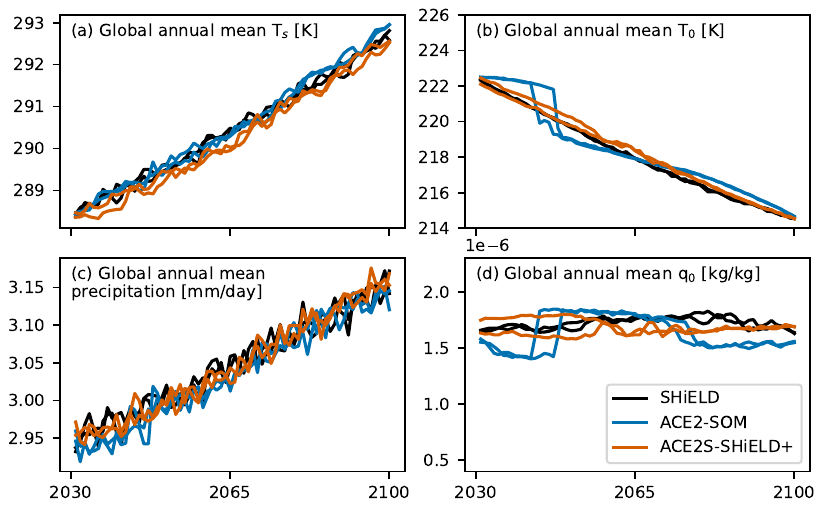}
\caption{Global annual mean time series of surface temperature (a), stratospheric temperature (b), precipitation rate (c), and stratospheric specific total water (d) in a two-member ensemble of \num{70}-year SOM-coupled 2pctCO$_2$ simulations with SHiELD, ACE2-SOM, and ACE2S-SHiELD+.}
\label{fig:2pctCO2}
\end{figure}

ACE2S-SHiELD+ also exhibits comparable skill to ACE2-SOM in SOM-coupled inference with CO$_2$ increasing at a rate of \SI{2}{\percent \per \year}.  Figure~\ref{fig:2pctCO2} shows the global annual mean surface temperature, stratospheric temperature, precipitation rate, and stratospheric specific total water in a two-member ensemble of 2pctCO$_2$ simulations with SHiELD, ACE2-SOM, and ACE2S-SHiELD+.  We find that ACE2S-SHiELD+ exhibits comparable skill for tropospheric and boundary flux variables, with less extreme regime shifts in the stratosphere.

\subsection{Prescribed SST test cases with decoupled SST and CO$_2$}

We now transition to out-of-sample forcing scenarios where ACE2S-SHiELD+ improves meaningfully over the baseline ACE2-SHiELD and ACE2-SOM models.  With a prescribed SST it is possible to vary the SST and CO$_2$ concentration independently from each other and therefore there are a number of potential experiments.  We will discuss results from experiments starting from the relatively mild to the more extreme.

\subsubsection{AMIP constant CO$_2$ inference}

\begin{figure}
\noindent\includegraphics[width=\textwidth]{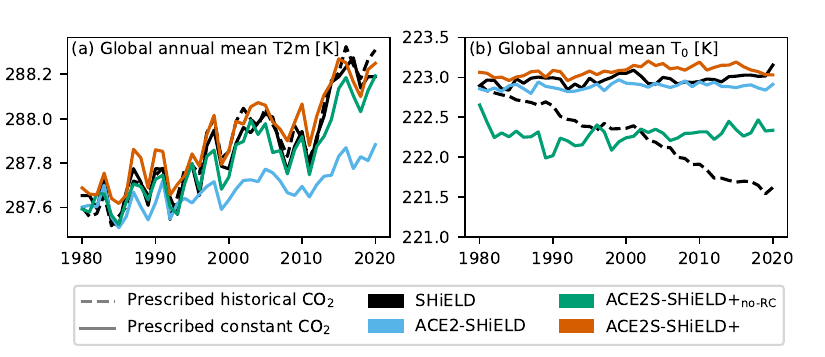}
\caption{AMIP inference initialized in 1979 with fixed CO$_2$ (solid lines): global annual mean \SI{2}{\m} temperature (a) and stratospheric temperature (b) in SHiELD, ACE2-SHiELD, ACE2S-SHiELD+$_{\text{no-RC}}$, and ACE2S-SHiELD+.  To give a sense for the impact of holding the CO$_2$ constant on each of these fields, the dashed black line shows their evolution in SHiELD in a traditional AMIP simulation with time-varying CO$_2$.}
\label{fig:amip-constant-CO2}
\end{figure}

\citeA{Wat2025} noted that a simple example that illustrates the incorrect separation of SST and CO$_2$ sensitivity in ACE2-SHiELD is AMIP inference with CO$_2$ held constant.  Figure~\ref{fig:amip-constant-CO2} shows the global annual mean time series of two meter and stratospheric temperature in AMIP inference initialized in 1979 where CO$_2$ is held fixed at the 1979 level.  ACE2-SHiELD places too much weight on the CO$_2$ concentration relative to SST in setting the \SI{2}{\m} temperature trend, indicated by the fact that global annual mean \SI{2}{\m} temperature increases by only about \SI{0.3}{\K} between 1980 and 2020; despite this, it accurately predicts that the stratospheric temperature should remain roughly constant at the 1979 level.  ACE2S-SHiELD+ on the other hand accurately emulates both the near-surface and stratospheric temperature under these conditions.  Interestingly, with the exception of a cold bias in the stratospheric temperature, ACE2S-SHiELD+ trained without random-CO$_2$ data does a reasonable job in this test case, though we will show it can break down in more extreme forcing scenarios.  Training with or without energy conservation does not make a robust qualitative difference (not shown).

\subsubsection{AMIP +4 K inference}

\begin{figure}
\noindent\includegraphics[width=\textwidth]{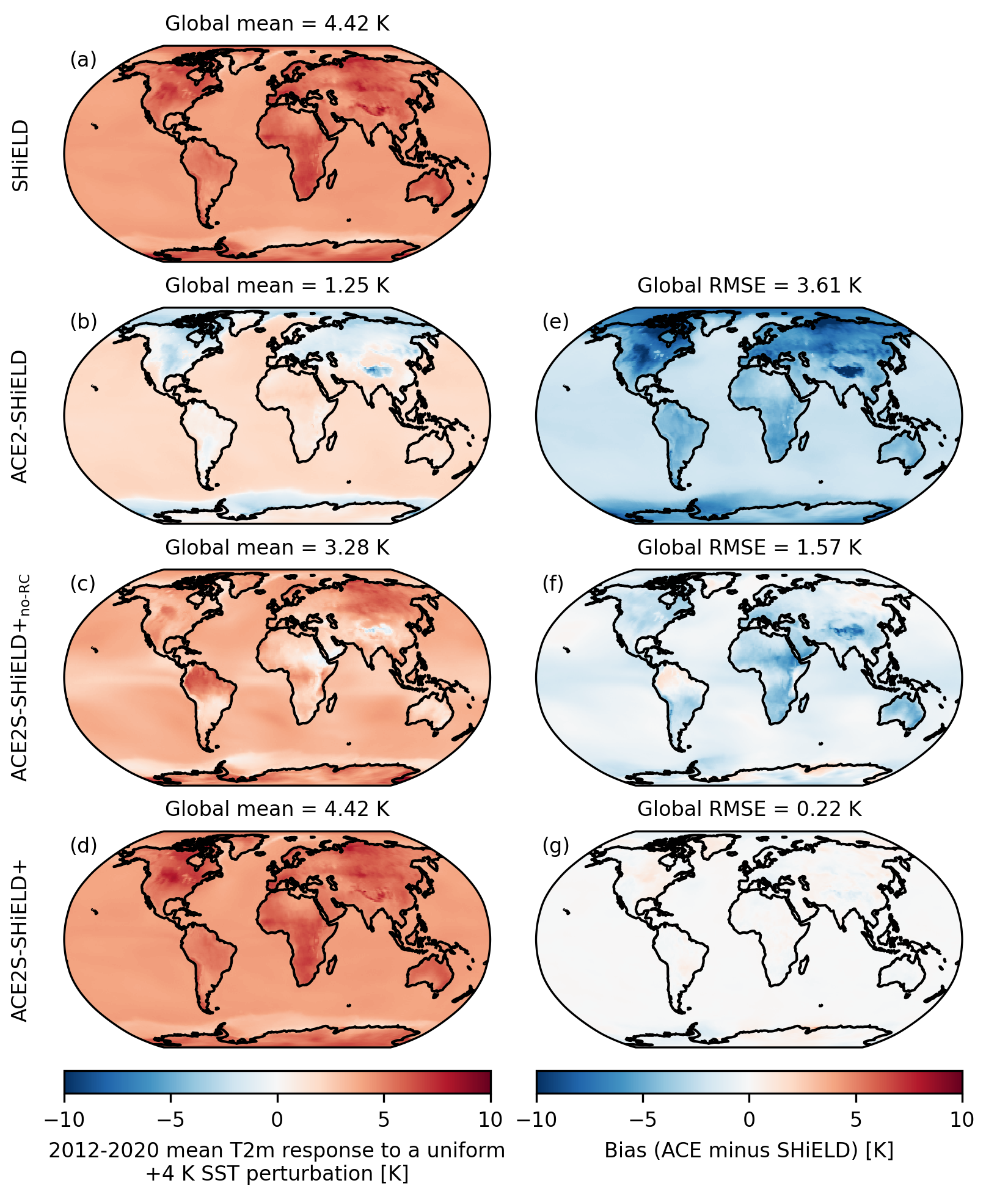}
\caption{2012 through 2020 mean difference in \SI{2}{\m} temperature between an AMIP \SI[retain-explicit-plus]{+4}{\K} simulation and an AMIP simulation with SHiELD (a), ACE2-SHiELD (b), ACE2S-SHiELD+$_{\text{no-RC}}$ (c), and ACE2S-SHiELD+ (d).  Panels (e)-(g) show the response pattern error relative to the target SHiELD.}
\label{fig:amip-p4K}
\end{figure}

A more extreme idealized climate change experiment used to study climate sensitivity \cite<e.g.,>{Ces1988, Mer2024} and cloud feedbacks \cite{Web2017} is an AMIP simulation with sea surface temperatures increased uniformly by \SI[retain-explicit-plus]{+2}{\K} or \SI[retain-explicit-plus]{+4}{\K}, but CO$_2$ remaining at historical levels.  ACE2-SHiELD (and ACE2-ERA5), trained only on data from the historical period, have well-documented spurious behavior in this context, for instance with \SI{2}{\m} temperature increasing only slightly over ocean and decreasing over land \cite{Wat2025} and a muted temperature response throughout the depth of the atmosphere despite a surprisingly realistic precipitation response \cite{Zha2026}.  

Figure~\ref{fig:amip-p4K} shows the \SI{2}{\m} temperature response in an AMIP \SI[retain-explicit-plus]{+4}{\K} experiment in SHiELD, ACE2-SHiELD, ACE2S-SHiELD+$_{\text{no-RC}}$, and ACE2S-SHiELD+.  Here the response is shown averaged only over the held out period of the standard AMIP simulations used in training, 2012 through 2020.  As described in previous studies, ACE2 trained on only AMIP-style data predicts cooling over land and sea-ice regions and muted warming over ocean.  ACE2S-SHiELD+ predicts a highly accurate response, with global RMSE of only \SI{0.22}{\K} relative to the target.  Looking at the responses of other fields, like outgoing longwave radiation (Figure~S4), upward shortwave radiative flux at the top of the atmosphere (Figure~S5), vertically resolved temperature (Figure~S6) and zonal wind (Figure~S7), or the \num{99.9}th percentile of daily-mean precipitation rate (Figure~S8), we see similarly strong agreement.  While training on a mixture of AMIP and equilibrium-climate data, and using the modified architecture and loss function of ACE2S helps some, training with random-CO$_2$ data, which separates SST perturbations from CO$_2$ changes, is key to obtaining accurate results.  Similar to the AMIP constant CO$_2$ test case, training with or without energy conservation does not make a robust qualitative difference (not shown).

\subsubsection{Ramped-SST-random-CO$_2$ inference}

\begin{figure}
\noindent\includegraphics[width=\textwidth]{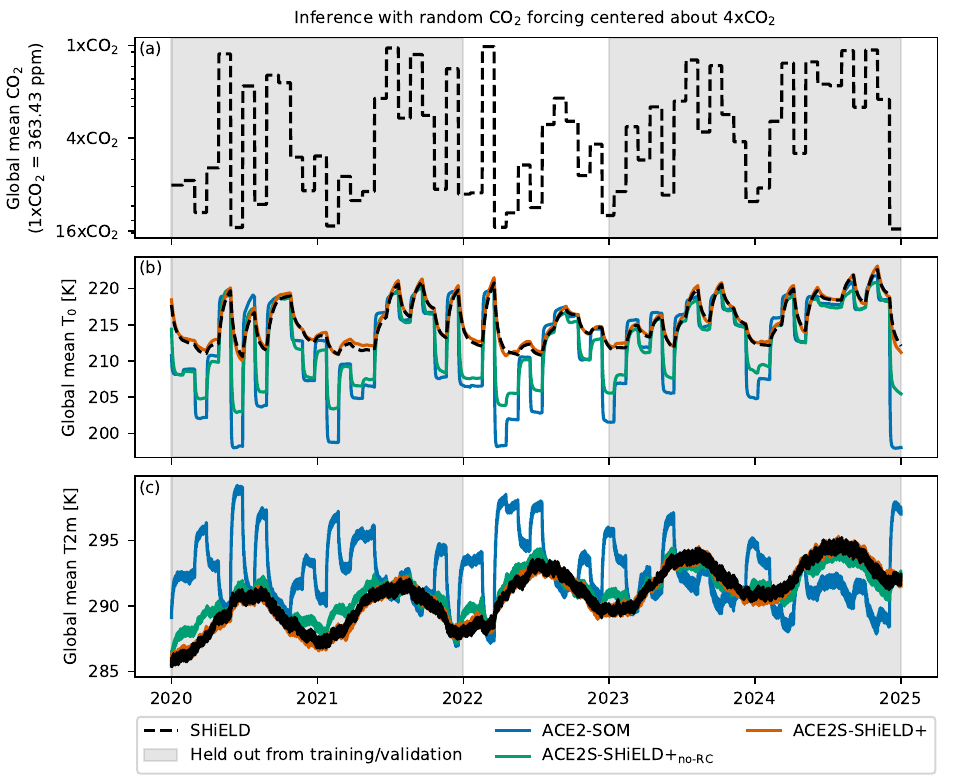}
\caption{CO$_2$ forcing (a), global mean stratospheric temperature (b), and global mean \SI{2}{\m} temperature (c) in mostly held-out random-CO$_2$ inference centered about 4xCO$_2$ in SHiELD, ACE2-SOM, ACE2S-SHiELD+$_{\text{no-RC}}$, and ACE2S-SHiELD+.  Periods where the CO$_2$ forcing was fully held out from training or validation are shaded in gray.  Note we have inverted the y-axis in panel (a) to highlight the expected inverse relationship between CO$_2$ concentration and the global mean stratospheric temperature in panel (b).}
\label{fig:random-CO2}
\end{figure}

We have held out four out of five years from the last random-CO$_2$ ensemble member runs centered about 1x, 2x, and 4xCO$_2$ completely from training and validation, and so they can be used as a final challenging data-ocean test case.  Figure~\ref{fig:random-CO2} shows the 6-hourly time series of CO$_2$ used as a forcing, global mean stratospheric temperature, and global mean \SI{2}{\m} temperature in ramped-SST-random-CO$_2$ inference centered about 4xCO$_2$ in SHiELD, ACE2-SOM, ACE2S-SHiELD+$_{\text{no-RC}}$, and ACE2S-SHiELD+.  In SHiELD it shows the tight inverse relationship between CO$_2$ concentration and stratospheric temperature, with low stratospheric temperature associated with high CO$_2$ concentrations and vice versa, and the fact that the \SI{2}{\m} temperature is largely modulated by the increasing SST forcing.  

Not surprisingly, including random-CO$_2$ data in training is necessary for accurate emulation of this case in the troposphere and stratosphere.  ACE2-SOM shows clear imprinting of the CO$_2$ concentration on the global mean \SI{2}{\m} temperature; while ACE2S-SHiELD+$_{\text{no-RC}}$ does meaningfully better, suggesting some improvement comes from the inclusion of AMIP data and the new architecture and loss function, it still exhibits larger biases than ACE2S-SHiELD+.  The cold biases in stratospheric temperature associated with models trained without random-CO$_2$ data can be attributed in part to the fact that they only saw CO$_2$ concentrations between 1x and 4xCO$_2$ during training, while models trained with it were exposed to CO$_2$ concentrations potentially as high as 16xCO$_2$.  Exposure to abrupt CO$_2$ changes during training also ensures emulating the correct timescale of the response.  In contrast, even when the CO$_2$ is within the range seen during training, in models trained without random-CO$_2$ data the stratospheric temperature adjusts almost immediately to the predicted equilibrium, producing a step-like time series in global mean $T_0$.  In reality it should relax to a new equilibrium with an $e$-folding timescale on the order of a week \cite{Blo2021, Mah2026}.

\subsection{Slab ocean test cases with strongly mismatched SST and CO$_2$}

When an atmosphere model is coupled to an interactive ocean, we can no longer directly control the SST.  The main way we can impose a mismatch between the CO$_2$ and the SST is to abruptly change the CO$_2$ associated with an equilibrium-climate initial condition.  The traditional example of this is the abrupt 4xCO$_2$ experiment of the Coupled Model Intercomparison Project DECK \cite{Eyr2016}.  The SST starts out cold relative to the CO$_2$ concentration and warms gradually to re-equilibrate following the large perturbation.  In this way, the early period of the abrupt 4xCO$_2$ case represents the reverse situation as the AMIP \SI[retain-explicit-plus]{+4}{\K} simulation, where the SST is warm relative to the CO$_2$ concentration throughout.  

\subsubsection{Abrupt 4xCO$_2$ inference ensemble}

\begin{figure}
\noindent\includegraphics[width=\textwidth]{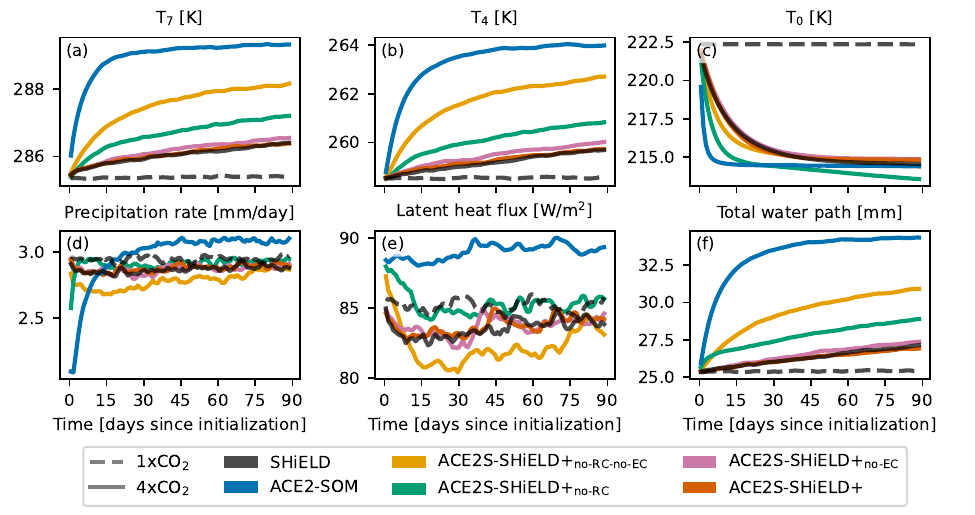}
\caption{Global, daily, and ensemble mean time series of lowest level temperature (a), mid-tropospheric temperature (b), stratospheric temperature (c), precipitation rate (d), latent heat flux (e), and total water path (f) in SOM-coupled abrupt 4xCO$_2$ simulations (solid lines) with SHiELD, ACE2-SOM, ACE2S-SHiELD+$_{\text{no-RC-no-EC}}$, ACE2S-SHiELD+$_{\text{no-RC}}$, ACE2S-SHiELD+$_{\text{no-EC}}$, and ACE2S-SHiELD+.  The dashed black line in each panel shows the ensemble-mean behavior of SHiELD with the same initial conditions, but without the abrupt CO$_2$ perturbation.}
\label{fig:abrupt-4xCO2-ensemble-temperature-hydrological-cycle}
\end{figure}

\begin{figure}
\noindent\includegraphics[width=\textwidth]{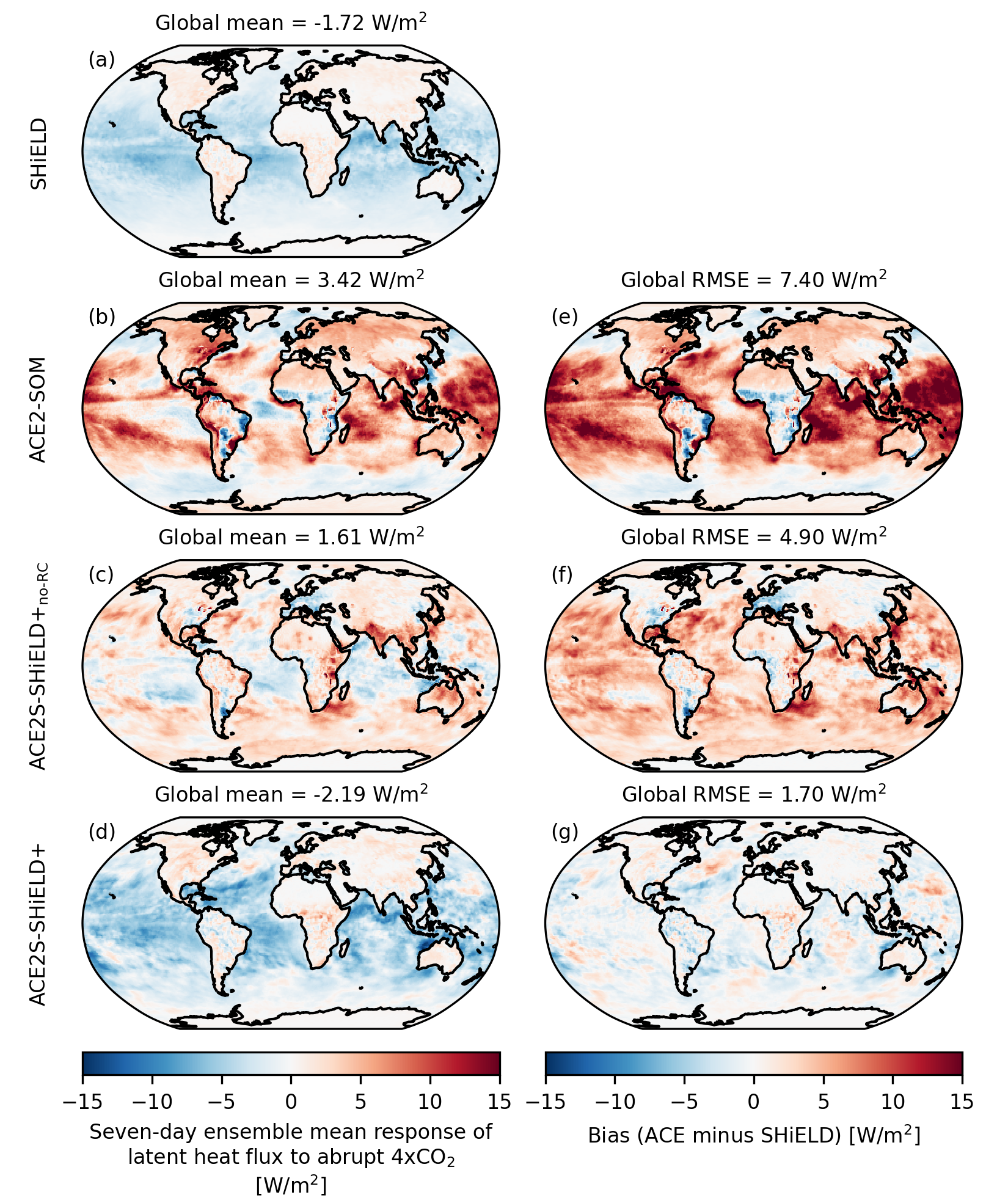}
\caption{Time-and-ensemble-mean response to abrupt 4xCO$_2$ of latent heat flux over the initial \num{7} days of SOM-coupled simulation with SHiELD (a), ACE2-SOM (b), ACE2S-SHiELD+${_\text{no-RC}}$ (c), and ACE2S-SHiELD+ (d). Panels (e)-(g) show the corresponding pattern errors of the ML models relative to SHiELD.}
\label{fig:abrupt-4xCO2-LHTFLsfc-spatial-pattern}
\end{figure}

We will begin by zooming in on the initial mean response in a \num{36}-member ensemble of \num{90}-day SOM-coupled abrupt 4xCO$_2$ simulations initialized a month apart over the course of three years.  This is a common method for robustly studying the radiative, cloud, and hydrological cycle responses to abrupt 4xCO$_2$, since it smooths out weather noise and samples responses throughout the annual cycle.  While these kinds of ensemble experiments are often run with prescribed SST to isolate the adjustment of the atmosphere to the change in CO$_2$, for simplicity we run in SOM-coupled mode, since we are ultimately interested in getting the coupled response correct, and results tend to be similar on these short timescales where the ocean temperature has little time to change \cite{Kam2013, Kam2015, Mah2026}.

\begin{figure}
\noindent\includegraphics[width=\textwidth]{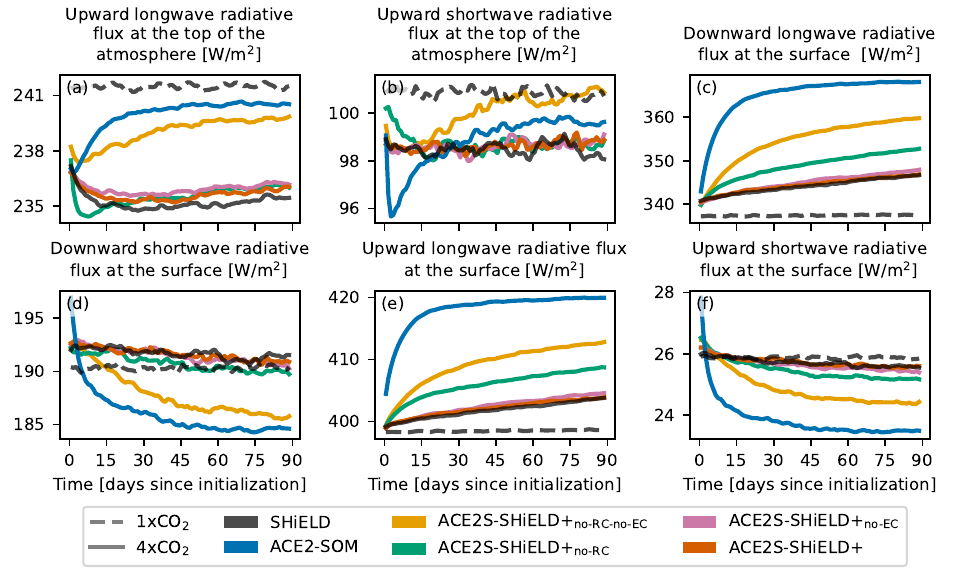}
\caption{As in Figure~\ref{fig:abrupt-4xCO2-ensemble-temperature-hydrological-cycle} but for the global, daily, and ensemble mean time series of upward longwave radiative flux at the top of the atmosphere (a), upward shortwave radiative flux at the top of the atmosphere (b), downward longwave radiative flux at the surface (c), downward shortwave radiative flux at the surface (d), upward longwave radiative flux at the surface (e), and upward shortwave radiative flux at the surface (f).}
\label{fig:abrupt-4xCO2-ensemble-radiation}
\end{figure}

Figure~\ref{fig:abrupt-4xCO2-ensemble-temperature-hydrological-cycle} shows the global, daily, and ensemble mean response of the temperature at different levels of the atmosphere (\ref{fig:abrupt-4xCO2-ensemble-temperature-hydrological-cycle}a through \ref{fig:abrupt-4xCO2-ensemble-temperature-hydrological-cycle}c) and the hydrological cycle (\ref{fig:abrupt-4xCO2-ensemble-temperature-hydrological-cycle}d through \ref{fig:abrupt-4xCO2-ensemble-temperature-hydrological-cycle}f) with SHiELD, ACE2-SOM, and the four configurations of ACE2S-SHiELD+ trained for this study.  The expected temperature response from SHiELD is a gradual warming in the troposphere at a rate of about \SI{0.25}{\K \per \year} and rapid cooling in the stratosphere from about \SI{222.5}{\K} to \SI{215}{\K} after about \num{60} days.  As discussed in \citeA{Cla2025b}, the state in ACE2-SOM shifts almost immediately to that of a 4xCO$_2$ equilibrium climate, fully equilibrating in roughly \num{90} days in the troposphere and less than \num{15} days in the stratosphere.  Switching to the new architecture and loss function of ACE2S and training on a mixture of AMIP and equilibrium-climate data (ACE2S-SHiELD+$_{\text{no-RC-no-EC}}$), and further imposing global energy conservation (ACE2S-SHiELD+$_{\text{no-RC}}$) help slow the response, but not to the point of matching the timescales of SHiELD.  It is only when we include random-CO$_2$ data in training that we are able to accurately emulate the evolution of the temperature at all levels of the atmosphere.  Once random-CO$_2$ data is included, imposing energy conservation is no longer critical, since ACE2S-SHiELD+$_{\text{no-EC}}$ manages to approximately conserve energy even in the absence of the constraint.

\begin{figure}
\noindent\includegraphics[width=\textwidth]{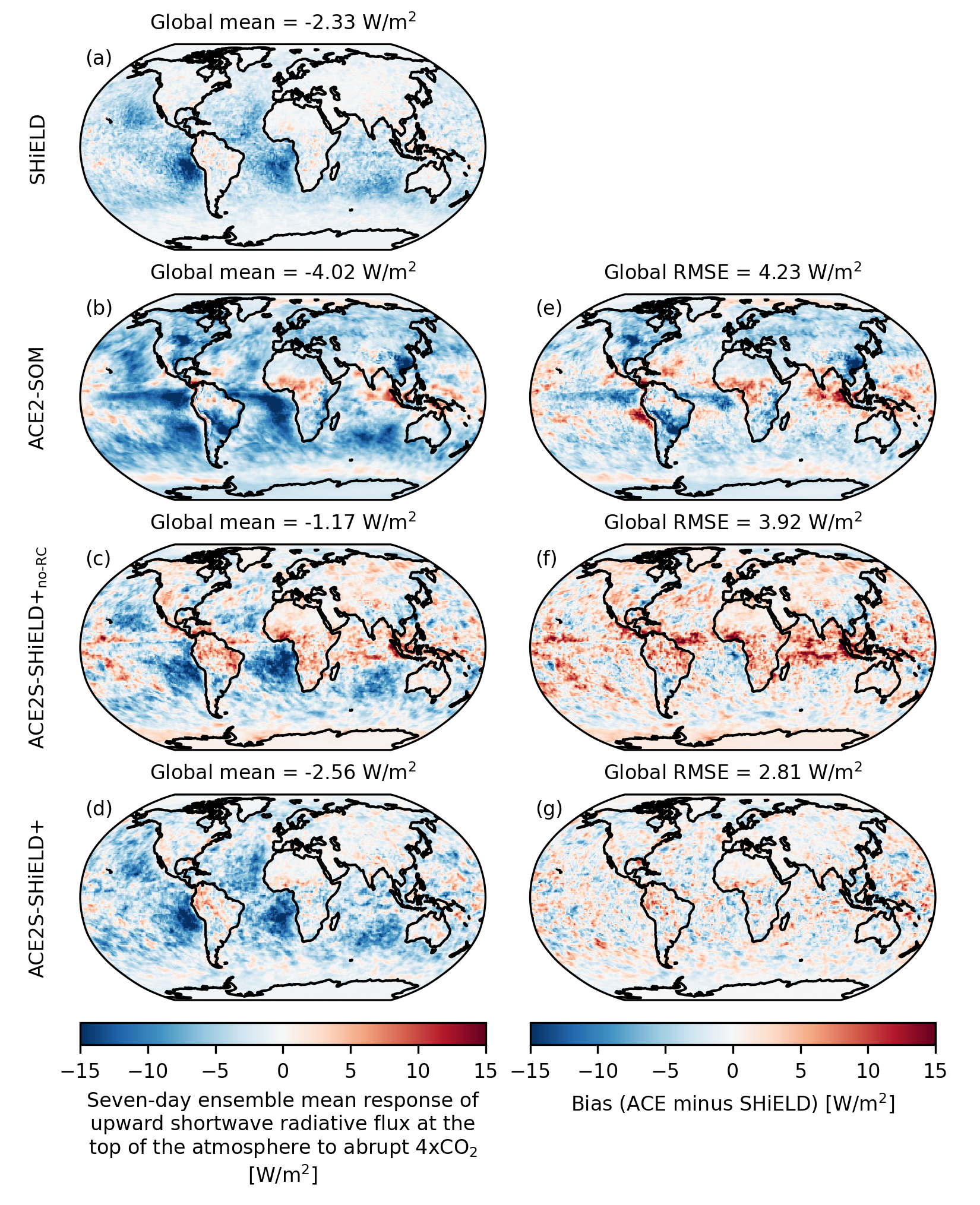}
\caption{As in Figure~\ref{fig:abrupt-4xCO2-LHTFLsfc-spatial-pattern} but for upward shortwave radiative flux at the top of the atmosphere.}
\label{fig:abrupt-4xCO2-USWRFtoa-spatial-pattern}
\end{figure}

In terms of the hydrological cycle, ACE2-SOM similarly responds too quickly, with the latent heat flux immediately taking on its typical value in a 4xCO$_2$ equilibrium climate, and the global mean total water path increasing by almost \SI{10}{\mm} in \num{90} days.  At the beginning of abrupt 4xCO$_2$ inference, the precipitation decreases sharply in ACE2-SOM by almost \SI{1}{\mm \per \day}.  While there are physical reasons we would expect the precipitation to decrease in this scenario, this is roughly an order of magnitude more than that exhibited in SHiELD or CMIP6 models \cite{Mah2026}.  Rather than an exaggerated representation of an otherwise physically correct response, the decrease in the case of ACE2-SOM can mainly be explained by the action of the moisture corrector scaling the precipitation down to maintain a closed global moisture budget in the presence of a large total water path tendency and a comparatively modest increase in the latent heat flux.  The new models trained for this study do better.  The models trained without random-CO$_2$ data get the right sign of the response in precipitation and latent heat flux, though still exhibit moisture-corrector-related pathologies due to the overly rapid increase in total water path.  Models trained with random-CO$_2$ data, on the other hand, match the response of SHiELD quite well, not just in the global mean, but also in the spatial pattern.

Following \citeA{Mah2026} we can look at the spatial pattern of the response of different variables averaged over the first \num{7} days and across all ensemble members.  Here the response is defined as the difference between rollouts starting from the same \num{36} initial conditions, with and without abruptly quadrupling CO$_2$.  The target latent heat flux response is shown in Figure~\ref{fig:abrupt-4xCO2-LHTFLsfc-spatial-pattern}a.  It is characterized by a relatively smooth decrease over the oceans, which is largest in the subtropics and decreases towards the high latitudes, and a slight increase over land regions.  This is comparable to that observed by \citeA{Mah2026} in E3SM.  

For comparison Figures~\ref{fig:abrupt-4xCO2-LHTFLsfc-spatial-pattern}b through \ref{fig:abrupt-4xCO2-LHTFLsfc-spatial-pattern}d show the responses as simulated by ACE2-SOM, ACE2S-SHiELD+$_{\text{no-RC}}$, and ACE2S-SHiELD+; the panels on the right show pattern errors relative to SHiELD.  ACE2-SOM and ACE2S-SHiELD+$_{\text{no-RC}}$ exhibit similar response patterns, with increases over most of the oceans.  Though slightly noisier, ACE2S-SHiELD+ data responds quite similarly to SHiELD, roughly getting the correct land-ocean contrast and global mean.  If we look at the response pattern of precipitation (Figure~S9), we find a similar story.  There the expected response is highly concentrated in the deep tropics around the intertropical convergence zone (ITCZ).  While a fair bit noisier in the subtropics and mid-latitudes, ACE2S-SHiELD+ produces roughly the correct spatial pattern and global mean.  Results without energy conservation are qualitatively similar (not shown).

An abrupt change in CO$_2$ naturally has direct and indirect effects on radiation.  The global, daily, and ensemble mean evolution of the surface and top of atmosphere radiative fluxes is shown in Figure~\ref{fig:abrupt-4xCO2-ensemble-radiation}.  The outgoing longwave radiative flux decreases immediately as a direct consequence of the increase in CO$_2$, and further decreases over the first month until it starts slowly increasing as the surface and troposphere warm.  While ACE2-SOM roughly captures the immediate decrease, the outgoing longwave flux increases nearly back to its equilibrium value within the first month.  Models trained with random-CO$_2$ data, whether we impose global energy conservation or not, match this evolution of SHiELD well.  They also capture well the responses of the longwave radiative fluxes at the surface.  The downward longwave radiative flux increases immediately due to the greenhouse effect and then further increases as the troposphere warms and moistens, while upward longwave radiative flux increases gradually following the increase in the temperature of the slab ocean and land surface.  As shown in Figure~S10, the spatial pattern of the \num{7}-day ensemble mean response of the upward longwave radiative flux at the top of the atmosphere is characterized by a fairly smooth decrease over much of the globe, with the exception of a mild increase in land ice regions.  ACE2S-SHiELD+ and ACE2S-SHiELD+$_{\text{no-EC}}$ (not shown) accurately capture this pattern.

In terms of global mean shortwave radiation, the upward flux at the top of the atmosphere decreases and the downward flux at the surface increases with the quadrupling of the CO$_2$.  Similar to the longwave radiative fluxes, the versions of ACE that predict this most accurately are those that were trained including random-CO$_2$ data, though ACE2S-SHiELD+$_{\text{no-RC}}$roughly gets the correct qualitative picture.  While CO$_2$ is a mild absorber of shortwave radiation, which may contribute some to the decrease in the upward flux at the top of the atmosphere \cite{Pin2020}, these responses can primarily be explained by a decrease in clouds \cite{Zel2013}.  Clouds reflect a portion of shortwave radiation back to space before it can reach the ground, meaning a decrease in clouds results in less reflected shortwave radiation within the atmosphere and more reaching the surface.  Figure~\ref{fig:abrupt-4xCO2-USWRFtoa-spatial-pattern}a illustrates the spatial pattern of the \num{7}-day ensemble mean response in upward shortwave radiative flux at the top of the atmosphere in SHiELD.  There is a decrease over the oceans in nearly all but the polar regions, which is largest in areas where marine stratocumulus clouds are common, like off the west coasts of Africa, the Americas, and Australia.  Conversely, there is little change over land.  Despite not explicitly emulating clouds, ACE2S-SHiELD+ accurately captures this response, as shown in Figure~\ref{fig:abrupt-4xCO2-USWRFtoa-spatial-pattern}d and \ref{fig:abrupt-4xCO2-USWRFtoa-spatial-pattern}g.

Taken together, these results, particularly those in Figure~\ref{fig:abrupt-4xCO2-ensemble-radiation}, demonstrate that ACE2S-SHiELD+ accurately emulates the global mean evolution of all the surface and top of atmosphere radiative fluxes, an important test of the physicality of its response to the radiative forcing introduced by abruptly quadrupling CO$_2$.  ACE2S-SHiELD+$_{\text{no-RC-no-EC}}$ and ACE2S-SHiELD+$_{\text{no-RC}}$ on the other hand perform worse in this regard in part due to biases in the response of the prognostic state and the implicit representation of the response of clouds.

\subsubsection{10-year abrupt 4xCO$_2$ inference}

\begin{figure}
\noindent\includegraphics[width=\textwidth]{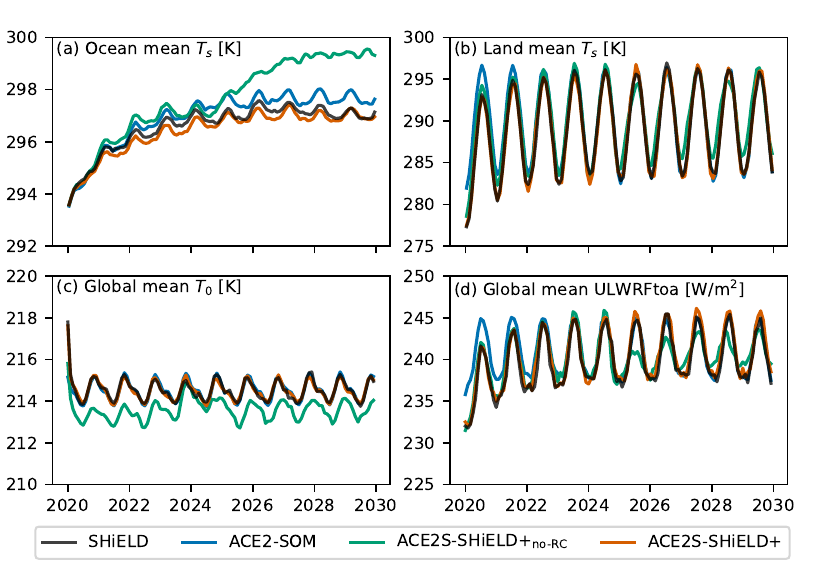}
\caption{Regional-mean evolution of various fields in a single \num{10}-year SOM-coupled abrupt 4xCO$_2$ inference simulation with SHiELD, ACE2-SOM, ACE2S-SHiELD+$_{\text{no-RC}}$, and ACE2S-SHiELD+.  Panels (a) and (b) show the ocean- and land-mean surface temperature, respectively; panel (c) shows the global mean uppermost model layer temperature; and panel (d) shows the global mean upward longwave radiative flux at the top of the atmosphere.}
\label{fig:10-year-abrupt-4xCO2}
\end{figure}

ACE2S-SHiELD+ not only captures the short-term response to an abrupt quadrupling of CO$_2$, but it also is capable of accurately simulating the full equilibration process.  Figure~\ref{fig:10-year-abrupt-4xCO2} shows the regional-mean evolution of a selected representative fields in a single \num{10}-year SOM-coupled abrupt 4xCO$_2$ simulation.  Of the models shown, only ACE2S-SHiELD+ obtains the expected evolution for ocean mean surface temperature, land mean surface temperature, stratospheric temperature, and upward longwave radiative flux at the top of the atmosphere.  ACE2-SOM, as shown previously, adjusts too quickly for all but the SOM-mediated surface temperature over the ocean, and while ACE2S-SHiELD+$_{\text{no-RC}}$ shows qualitatively better behavior in the beginning, it fails to find the correct equilibrium-climate attractor even after \num{10} years.

\section{Discussion and Conclusion}

In this study we have presented an efficient data-driven approach for training a machine-learning-based atmosphere model to respond accurately to independent SST and CO$_2$ forcings.  The approach relies on supplementing training data from more traditional types of reference simulations, like AMIP and equilibrium-climate slab ocean simulations with varying levels of CO$_2$, with data from simulations where the SST and CO$_2$ are varied strongly independently from each other.  With such a dataset, we trained a model we refer to as ACE2S-SHiELD+ that is more flexible than the previous baselines of ACE2-SHiELD and ACE2-SOM combined, with approximately \SI{25}{\percent} fewer samples than either were trained on alone.  We have shown that ACE2S-SHiELD+ accurately emulates SHiELD in both data and slab ocean configurations with a variety of SST and CO$_2$ forcings held out from training, addressing noted generalization issues to uniform SST perturbations \cite{Wat2025, Zha2026} and the transient response to an abrupt CO$_2$ change \cite{Cla2025b, Mah2026}.

ACE2S-SHiELD+ includes a constraint to ensure global energy conservation to within the average residual of its target physics-based model.  While the technical details differ slightly, this constraint is similar to that introduced in \citeA{Cha2025a} and \citeA{Sha2025}.  We find that in the absence of random-CO$_2$ data, including it helps improve the physicality of the response to an abrupt CO$_2$ change, but it is too crude to enable emulating it fully, since it does not guarantee accurate prediction of the boundary energy fluxes and applies a globally uniform temperature correction, which may not be representative of the precise spatial structure or form of the non-conservation error.  For the types of generalization we were interested in for this study, incorporating random-CO$_2$ data in training made the overwhelming difference. A more careful study training models with more random weight initializations would likely be necessary to assess whether including energy conservation led to a detectable improvement in emulating the time mean or other statistics of climate, though \citeA{Sha2025} noted that it can help slightly with medium range weather forecast skill.  However, the fact that it provides improved interpretability without degrading skill in our case is meaningful in itself.

While ACE2S-SHiELD+ increases the range of climate change experiments that can be accurately attempted with a single emulator, there are a number of opportunities for future research and development.  ACE2S-SHiELD+ remains limited in its representation of other Earth system components, like the ocean, land, and sea ice.  Though it can be coupled to a slab ocean, which allows for an interactive SST (albeit without circulation feedbacks), sea ice fraction is prescribed as an annual repeating climatology, limiting its sensitivity to CO$_2$ relative to more comprehensive models.  An important direction will be how one can extend this approach to coupled emulators, like SamudrACE \cite{Dun2026}.  Will it be sufficient in these contexts to ensure the atmosphere predicts the appropriate fluxes for coupling, or will similarly sophisticated training datasets be necessary for the other components?  Research suggests that at least for sea ice, models like FloeNet may be able to generalize if given the appropriate fluxes alone, but this has yet to be tested in a coupled setting \cite{Gre2026}.

Another important dimension is generalization to other types of forcings.  CO$_2$ is a single well-mixed greenhouse gas.  There are numerous other greenhouse gases, like CH$_4$ or N$_2$O, and aerosols, which are shorter lived and highly spatially heterogeneous, all of which can vary independently of CO$_2$ \cite{Ria2017, Mei2020}.  A machine-learning-based weather/climate model that could accurately respond to different combinations of these forcings would be needed to be able to make projections under a full range of diverse possible emissions scenarios; there is an extensive history of doing this with reduced complexity models and statistical approaches, though these are limited in the breadth and depth of climate information they can provide \cite{Nic2020, Teb2025a}. How to best develop appropriate training datasets to learn the response to these kinds of forcings is an open question, though it is possible some inspiration could be drawn from this work.

Lastly, the ability to generate training data with arbitrary forcings exists only in the perfect model emulation framework.  Whether it is possible to develop a fully machine-learning-based model that is more accurate than existing physics-based models during the observed period, but retains plausible responses in forcing scenarios like those described in this study is something that future work will need to address.  When trained only on ERA5 reanalysis and without CO$_2$ as a forcing, some machine-learning-based models, like ArchesWeatherGen-V2 \cite{Cou2026} or DLESyM \cite{Cre2025a}, exhibit more realistic responses to uniform SST perturbations than ACE2.1 trained in the same manner \cite{Hen2026}.  This suggests architecture may play a role in generalization ability, but does not necessarily give an indication of whether these models would better separate the effects of SST and CO$_2$ if trained with both as inputs.  This hypothesis also would benefit from clean testing in a perfect model emulation framework where the target is known.  Hybrid approaches like NeuralGCM \cite{Koc2024} or an approach like that in \citeA{Mah2026} may have some advantages in that their physics-based components will generalize well, though cleanly separating learned from physics-based processes can also be a challenge.

\appendix

\section{Global energy corrector}
\label{sec:global-energy-corrector}

Here we provide a detailed description of the global energy corrector in ACE.  It shares some aspects with the corrector introduced in \citeA{Cha2025a} and \citeA{Sha2025}, but also has some minor differences in its definition of total energy, inclusion of frozen precipitation in the budget, and form of correction.

A general definition of the total energy in an atmosphere model grid cell, similar to that described in \citeA{Zho2022a}, is given by:
\begin{equation}
e = c_{vm} T + L_{v}(T) q_v - L_{f}(T) q_s + gz + \frac{1}{2} \mathbf{v} \cdot \mathbf{v},
\end{equation}
where $c_{vm}$ is the specific heat of moist air at constant volume, $T$ is the temperature, $L_{v}(T)$ is the temperature-dependent latent heat of vaporization, $q_v$ is the water vapor mixing ratio, $L_{f}(T)$ is the temperature-dependent latent heat of fusion, $q_s$ is the mixing ratio of solid hydrometeors, $g$ is the gravitational acceleration (\SI{9.80665}{\m \per \s \squared}), $z$ is the geopotential height, and $\mathbf{v}$ is the three-dimensional vector wind.  Given the vertical coarsening employed by ACE, and its aggregation of all water species into a single specific total water tracer, it is not possible to compute this exactly.  We therefore make several pragmatic assumptions to arrive at a simpler definition of total energy:
\begin{itemize}
\item The specific heat of moist air at constant volume can be approximated by the specific heat of dry air at constant volume, $c_v$ (\SI{717.55}{\J \per \kg \per \K}).
\item The latent heat of vaporization can be approximated as a temperature-independent constant, $L_v$ (\SI{2.5e6}{\J \per \kg}).
\item The mixing ratio of solid hydrometeors is small and can be treated as zero.
\item The kinetic energy is small compared to the internal energy and moist terms, and can be treated as zero.
\end{itemize}
With these assumptions, we define the total energy in ACE as:
\begin{equation}
e_{ACE} = c_v T + L_v q + gz,
\end{equation}
where $q$ is the specific total water.  These assumptions are similar, but slightly simpler still than those of \citeA{Sha2025}, who retain the contribution of kinetic energy.  We find these assumptions to be reasonable, since we do not obtain a meaningfully smaller global energy budget residual in our target data with $e$ than with $e_{ACE}$.
% See this notebook for evidence for the above statement: https://github.com/ai2cm/explore2/blob/main/spencerc/2026-02-17-multi-CO2-ACE-SHiELD/2026-05-19-energy-budget-residual.ipynb

\begin{figure}
\begin{center}
\centering
\noindent\includegraphics[width=\textwidth]{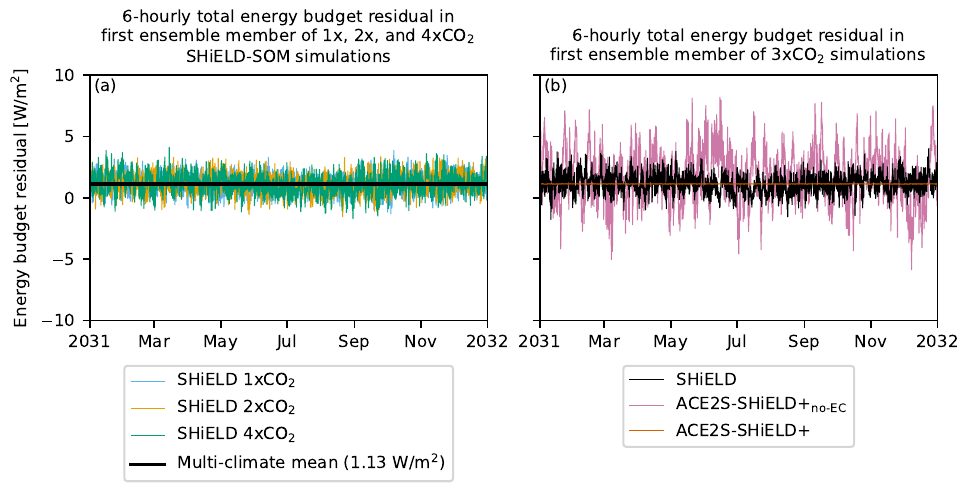}
\caption{6-hourly global total energy budget residual during the first post-spin-up year of the first ensemble member of the 1x, 2x, and 4xCO$_2$ equilibrium climate simulations with SHiELD (a) and the same in 3xCO$_2$ equilibrium climate simulations with SHiELD, ACE2S-SHiELD+ trained with random-CO$_2$ data but without energy conservation, and ACE2S-SHiELD+ trained with random-CO$_2$ data with energy conservation (b).  This is representative of the energy budget residual in subsequent years, the full 1x, 2x, and 4xCO$_2$ time mean of which is shown in the black line of panel (a).  We use this value as the prescribed unaccounted heating, $H_o$, in all models trained with energy conservation imposed.}
\label{fig:unaccounted-heating}
\end{center}
\end{figure}

SHiELD exhibits an approximately climate-and-time-invariant unaccounted heating of \SI{1.13}{\W \per \m \squared}, computed based on the mean energy budget residual in the first ensemble members in the equilibrium 1x, 2x, and 4xCO$_2$ climates (Figure~\ref{fig:unaccounted-heating}a).  To account for this energy-budget non-closure, we define the global total energy budget in ACE as:
\begin{equation}
\frac{\Delta \left\{ \left< e_{ACE} \right> \right\}}{\Delta t} = \left\{ \overline{F_{net}} \right\} + H_o.
\end{equation}
The $\Delta$ denotes a difference across a \num{6}-hour timestep, the curly braces denote an area-weighted mean, the angle brackets denote a mass-weighted vertical integral, i.e. 
\begin{equation}
\left< \left( \cdot \right) \right> = \frac{1}{g} \int_0^{p_s} \left( \cdot \right)\, dp,
\end{equation}
$\overline{F_{net}}$ represents the time-mean net energy input to the atmosphere due to known physical sources and sinks over the \num{6}-hour timestep $\Delta t$, and $H_o$ represents the assumed constant unaccounted heating.  Broken down into physical components, $F_{net}$ is defined as:
\begin{equation}
F_{net} = R_{toa} - R_{sfc} + LH + SH + L_f P_{frozen},
\end{equation}
where $R_{toa}$ and $R_{sfc}$ are the net downward radiative fluxes at the top of atmosphere and surface, $LH$ and $SH$ are the latent and sensible heat surface turbulent fluxes, $L_f$ is the latent heat of fusion (\SI{3.34e5}{\J \per \kg}), and $P_{frozen}$ is the total frozen precipitation rate.  

To enforce a closed global energy budget at the end of each \num{6}-hour timestep, we compute a globally constant temperature correction that ensures the total change in energy over the timestep matches the net energy input to the atmosphere plus the configured unaccounted heating.  Taking into account the impact on the geopotential, this uniform temperature correction can be computed as:
\begin{equation}
\label{eq:energy-corrector}
\delta T = \frac{\Delta t \left( \left\{ \overline{F_{net}} \right\} + H_o \right) - \Delta \left\{ \left< e_{ACE} \right> \right\}}{ \left\{ \frac{1}{g} \left( c_v p_s + \int_0^{p_s} \int_p^{p_s} \frac{\left( 1 + \left( \frac{R_v}{R_d} - 1 \right) q \right) R_d}{p'} dp' dp \right) \right\}},
\end{equation}
where $R_d$ is the gas constant for dry air (\SI{287.05}{\J \per \kg \K}), $R_v$ is the gas constant for water vapor (\SI{461.5}{\J \per \kg \K}), $p$ is the pressure, and $p_s$ is the surface pressure.  The corrected temperature prediction then becomes:
\begin{equation}
T' = T + \delta T.
\end{equation}
As an illustration, Figure~\ref{fig:unaccounted-heating}b shows the time series of the 6-hourly energy budget residual during one year of 3xCO$_2$ equilibrium-climate inference with ACE2S-SHiELD+ trained with random-CO$_2$ data with and without the energy corrector.  With the corrector the residual is constrained to always be \SI{1.13}{\W \per \meter \squared}, while without it the residual has roughly the same time mean, but fluctuates with a greater amplitude than it does in SHiELD.

\section{Detailed summary of data configurations}
\label{sec:data-configurations}

Here we provide a table summarizing the data used for computing normalization statistics, training, validation, and inline inference for the new models in this study.  As described in the main text, there are two data configurations.  The first includes only AMIP and equilibrium-climate data, which we refer to as the configuration ``without random-CO$_2$,'' while the second includes AMIP, equilibrium-climate, and random-CO$_2$ data, which we refer to as ``with random-CO$_2$.''  We consider the configuration including random-CO$_2$ data to be our main data configuration, so unless otherwise noted, it can be assumed to be the configuration that was used.  We append the subscript ``no-RC'' to the names of models trained without random-CO$_2$ data to distinguish them from models that included it. 

\begin{sidewaystable}
%TC:ignore
\caption{Summary of data used for training models.  While the equilibrium-climate reference data comes from SOM-coupled SHiELD simulations, for simplicity, training, validation, and inline inference are always done with prescribed SST for models trained in this study to align with the ocean configurations of the AMIP and random-CO$_2$ sources.}
\centering
\begin{tabular}{p{0.15\textwidth} | p{0.2\textwidth} | p{0.2\textwidth} | p{0.2\textwidth} 
| p{0.2\textwidth}}
\hline
 Name & Statistics (\num{30} years) & Training (\num{90} years) & Validation (\num{6} or \num{9} years) & Inline inference (\num{120} years) \\
\hline
  Without random-CO$_2$ (no-RC) &   
  Equilibrium (\num{30} years):
  \begin{itemize}
  \item 1xCO$_2$-IC1: 2031--2040
  \item 2xCO$_2$-IC1: 2031--2040
  \item 4xCO$_2$-IC1: 2031--2040
  \end{itemize} 
  & AMIP (\num{45} years): 
  \begin{itemize}
  \item IC1: 1979--2008
  \item IC2: 
  \begin{itemize} 
  \item 1981-07--1986-06
  \item 1991-07--1996-06
  \item 2001-07--2006-06
  \end{itemize}
  \end{itemize}
  Equilibrium (\num{45} years):
  \begin{itemize}
  \item 1xCO$_2$-IC1: 2031--2040
  \item 1xCO$_2$-IC2: 2032--2036
  \item 2xCO$_2$-IC1: 2031--2040
  \item 2xCO$_2$-IC2: 2032--2036
  \item 4xCO$_2$-IC1: 2031--2040
  \item 4xCO$_2$-IC2: 2032--2036
  \end{itemize}
  & AMIP (\num{3} years):
  \begin{itemize}
  \item IC1: 2009--2011
  \end{itemize}
  Equilibrium (\num{3} years):
  \begin{itemize}
  \item 1xCO$_2$-IC2: 2038
  \item 2xCO$_2$-IC2: 2038
  \item 4xCO$_2$-IC2: 2038
  \end{itemize}
  & Equilibrium (\num{120} years):
  \begin{itemize}
  \item 1xCO$_2$-IC3: 8x 5-years
  \item 2xCO$_2$-IC3: 8x 5-years
  \item 4xCO$_2$-IC3: 8x 5-years
  \end{itemize}
  \\
  \hline
  With random-CO$_2$ & 
  Random-CO$_2$ (\num{30} years):
  \begin{itemize}
  \item 1xCO$_2$-IC1: 2020--2024
  \item 1xCO$_2$-IC2: 2020--2024
  \item 2xCO$_2$-IC1: 2020--2024
  \item 2xCO$_2$-IC2: 2020--2024
  \item 4xCO$_2$-IC1: 2020--2024
  \item 4xCO$_2$-IC2: 2020--2024
  \end{itemize}
  & AMIP (\num{30} years): 
  \begin{itemize}
  \item IC1: 1979--2008
  \end{itemize}
  Equilibrium (\num{30} years):
  \begin{itemize}
  \item 1xCO$_2$-IC1: 2031--2040
  \item 2xCO$_2$-IC1: 2031--2040
  \item 4xCO$_2$-IC1: 2031--2040
  \end{itemize}
  Random-CO$_2$ (\num{30} years):
  \begin{itemize}
  \item 1xCO$_2$-IC1: 2020--2024
  \item 1xCO$_2$-IC2: 2020--2024
  \item 2xCO$_2$-IC1: 2020--2024
  \item 2xCO$_2$-IC2: 2020--2024
  \item 4xCO$_2$-IC1: 2020--2024
  \item 4xCO$_2$-IC2: 2020--2024
  \end{itemize}
  & AMIP (\num{3} years):
  \begin{itemize}
  \item IC1: 2009--2011
  \end{itemize}
  Equilibrium (\num{3} years):
  \begin{itemize}
  \item 1xCO$_2$-IC2: 2038
  \item 2xCO$_2$-IC2: 2038
  \item 4xCO$_2$-IC2: 2038
  \end{itemize}
  Random-CO$_2$ (\num{3} years):
  \begin{itemize}
  \item 1xCO$_2$-IC3: 2022
  \item 2xCO$_2$-IC3: 2022
  \item 4xCO$_2$-IC3: 2022
  \end{itemize}
  & 
  Equilibrium (\num{120} years):
  \begin{itemize}
  \item 1xCO$_2$-IC3: 8x 5-years
  \item 2xCO$_2$-IC3: 8x 5-years
  \item 4xCO$_2$-IC3: 8x 5-years
  \end{itemize}
  \\
\hline
\multicolumn{5}{p{22cm}}{$^{a}$Time bounds are inclusive and represent the largest possible range of times that can be represented by dates that start with the provided bounds, i.e. 1981-07--1986-06 represents the range 1981-07-01T00:00:00--1986-06-30T18:00:00.}
\label{tab:data-configurations}
\end{tabular}
%TC:endignore
\end{sidewaystable}

\clearpage

\section*{Open Research Section}
The code for ACE is openly developed and can be found on GitHub at \url{https://github.com/ai2cm/ace}; the latest release, \texttt{v2026.5.1}, is archived on Zenodo \cite{McG2026}.  The ACE2S-SHiELD+ checkpoint featured in this study is available on Hugging Face at \url{https://huggingface.co/allenai/ACE2S-SHiELD-plus}, along with some sample initial conditions and forcing data.  The configuration information and code required for generating the data, running the experiments, and producing the figures for this study is organized in a GitHub repository located at \url{https://github.com/ai2cm/ace2s-shield-plus-paper}.  Once the manuscript is in final form, this paper repository will be archived on Zenodo and the full processed SHiELD datasets used for training and evaluation will be hosted publicly.  

The text in this manuscript was human-written.  AI tools, such as Claude Sonnet 4.6 and Claude Opus 4.6, were used for code generation in the development of ACE.  The Cursor IDE provided AI-based code completion and limited-scope code generation when writing experiment submission scripts and producing figures for this specific study.

\section*{Conflict of Interest disclosure}
The authors declare there are no conflicts of interest for this manuscript.

\acknowledgments

Ai2 is supported by the estate of Paul G. Allen.  The reference SHiELD simulations and portions of ACE training were completed with computational resources provided by the NOAA/Geophysical Fluid Dynamics Laboratory.  The remaining portions of ACE training and all evaluation were completed with Ai2 computational resources.  We thank Bill Collins and Ankur Mahesh for helpful discussions, which informed our approach to evaluating abrupt 4xCO$_2$ inference skill, and Bosong Zhang for constructive comments on an earlier version of the manuscript.

%%%%%%%%%%%%%%%%%%%%%%%%%%%%%%%%%%%%%%%%%%%%%%%
% REFERENCES and BIBLIOGRAPHY
%
%TC:ignore
\bibliography{references}

@article{Blo2021,
  title = {Radiative {{Relaxation Time Scales Quantified}} from {{Sudden Stratospheric Warmings}}},
  author = {Bloxam, Kevin and Huang, Yi},
  year = 2021,
  month = jan,
  journal = {Journal of the Atmospheric Sciences},
  volume = {78},
  number = {1},
  pages = {269--286},
  publisher = {American Meteorological Society},
  issn = {0022-4928, 1520-0469},
  doi = {10.1175/JAS-D-20-0015.1},
  urldate = {2026-06-04},
  chapter = {Journal of the Atmospheric Sciences},
  langid = {english}
}

@misc{Bon2023,
  title = {Spherical {{Fourier Neural Operators}}: {{Learning Stable Dynamics}} on the {{Sphere}}},
  shorttitle = {Spherical {{Fourier Neural Operators}}},
  author = {Bonev, Boris and Kurth, Thorsten and Hundt, Christian and Pathak, Jaideep and Baust, Maximilian and Kashinath, Karthik and Anandkumar, Anima},
  year = 2023,
  month = jun,
  eprint = {2306.03838},
  primaryclass = {cs},
  publisher = {arXiv},
  doi = {10.48550/arXiv.2306.03838},
  urldate = {2025-05-17},
  archiveprefix = {arXiv}
}

@misc{Bon2025,
  title = {{{FourCastNet}} 3: {{A}} geometric approach to probabilistic machine-learning weather forecasting at scale},
  shorttitle = {{{FourCastNet}} 3},
  author = {Bonev, Boris and Kurth, Thorsten and Mahesh, Ankur and Bisson, Mauro and Kossaifi, Jean and Kashinath, Karthik and Anandkumar, Anima and Collins, William D. and Pritchard, Michael S. and Keller, Alexander},
  year = 2025,
  month = jul,
  publisher = {arXiv},
  doi = {10.48550/arXiv.2507.12144},
  urldate = {2026-04-01},
  langid = {english}
}

@article{Ces1988,
  title = {A methodology for understanding and intercomparing atmospheric climate feedback processes in general circulation models},
  author = {Cess, Robert D. and Potter, Gerald L.},
  year = 1988,
  journal = {Journal of Geophysical Research: Atmospheres},
  volume = {93},
  number = {D7},
  pages = {8305--8314},
  issn = {2156-2202},
  doi = {10.1029/JD093iD07p08305},
  urldate = {2026-04-19},
  copyright = {Copyright 1988 by the American Geophysical Union.},
  langid = {english}
}

@misc{Cha2025a,
  title = {{{CAMulator}}: {{Fast Emulation}} of the {{Community Atmosphere Model}}},
  shorttitle = {{{CAMulator}}},
  author = {Chapman, William E. and Schreck, John S. and Sha, Yingkai and Gagne, David John and Kimpara, Dhamma and Zanna, Laure and Mayer, Kirsten J. and Berner, Judith},
  year = 2025,
  month = apr,
  eprint = {2504.06007},
  primaryclass = {physics.ao-ph},
  publisher = {arXiv},
  doi = {10.48550/arXiv.2504.06007},
  urldate = {2026-06-04},
  archiveprefix = {arXiv}
}

@article{Che2022,
  title = {Impact of {{Warmer Sea Surface Temperature}} on the {{Global Pattern}} of {{Intense Convection}}: {{Insights From}} a {{Global Storm Resolving Model}}},
  shorttitle = {Impact of {{Warmer Sea Surface Temperature}} on the {{Global Pattern}} of {{Intense Convection}}},
  author = {Cheng, Kai-Yuan and Harris, Lucas and Bretherton, Christopher and Merlis, Timothy M. and Bolot, Maximilien and Zhou, Linjiong and Kaltenbaugh, Alex and Clark, Spencer and Fueglistaler, Stephan},
  year = 2022,
  month = aug,
  journal = {Geophysical Research Letters},
  volume = {49},
  number = {16},
  pages = {e2022GL099796},
  issn = {1944-8007},
  doi = {10.1029/2022GL099796},
  urldate = {2022-09-03},
  langid = {english}
}

@article{Chi2026,
  title = {Modulation of {{Tropical Cyclogenesis}} by the {{Convectively Coupled Kelvin Waves}}: {{Insights From Data-Driven Climate Emulator ACE2}}},
  shorttitle = {Modulation of {{Tropical Cyclogenesis}} by the {{Convectively Coupled Kelvin Waves}}},
  author = {Chien, Mu-Ting and Barnes, Elizabeth A. and Maloney, Eric D.},
  year = 2026,
  journal = {Geophysical Research Letters},
  volume = {53},
  number = {2},
  pages = {e2025GL117387},
  issn = {1944-8007},
  doi = {10.1029/2025GL117387},
  urldate = {2026-03-27},
  copyright = {\copyright{} 2026. The Author(s).},
  langid = {english}
}

@article{Cla2025b,
  title = {{{ACE2-SOM}}: {{Coupling}} an {{ML Atmospheric Emulator}} to a {{Slab Ocean}} and {{Learning}} the {{Sensitivity}} of {{Climate}} to {{Changed CO2}}},
  shorttitle = {{{ACE2-SOM}}},
  author = {Clark, Spencer K. and {Watt-Meyer}, Oliver and Kwa, Anna and McGibbon, Jeremy and Henn, Brian and Perkins, W. Andre and Wu, Elynn and Harris, Lucas M. and Bretherton, Christopher S.},
  year = 2025,
  journal = {Journal of Geophysical Research: Machine Learning and Computation},
  volume = {2},
  number = {4},
  pages = {e2024JH000575},
  issn = {2993-5210},
  doi = {10.1029/2024JH000575},
  urldate = {2026-03-27},
  copyright = {\copyright{} 2025. The Author(s). Journal of Geophysical Research: Machine Learning and Computation published by Wiley Periodicals LLC on behalf of American Geophysical Union.},
  langid = {english}
}

@article{Cou2026,
  title = {{{ArchesWeatherGen}}: {{Skillful}} and compute-efficient probabilistic weather forecasting with machine learning},
  shorttitle = {{{ArchesWeatherGen}}},
  author = {Couairon, Guillaume and Singh, Renu and Charantonis, Anastase and Lessig, Christian and Monteleoni, Claire},
  year = 2026,
  month = apr,
  journal = {Science Advances},
  volume = {12},
  number = {17},
  pages = {eadx2372},
  publisher = {American Association for the Advancement of Science},
  doi = {10.1126/sciadv.adx2372},
  urldate = {2026-05-01}
}

@article{Cre2025a,
  title = {A {{Deep Learning Earth System Model}} for {{Efficient Simulation}} of the {{Observed Climate}}},
  author = {{Cresswell-Clay}, Nathaniel and Liu, Bowen and Durran, Dale R. and Liu, Zihui and Espinosa, Zachary I. and Moreno, Raul A. and Karlbauer, Matthias},
  year = 2025,
  journal = {AGU Advances},
  volume = {6},
  number = {4},
  pages = {e2025AV001706},
  issn = {2576-604X},
  doi = {10.1029/2025AV001706},
  urldate = {2026-03-27},
  copyright = {\copyright{} 2025. The Author(s).},
  langid = {english}
}

@article{Dun2024,
  title = {Application of the {{AI2 Climate Emulator}} to {{E3SMv2}}'s {{Global Atmosphere Model}}, {{With}} a {{Focus}} on {{Precipitation Fidelity}}},
  author = {Duncan, James P. C. and Wu, Elynn and Golaz, Jean-Christophe and Caldwell, Peter M. and {Watt-Meyer}, Oliver and Clark, Spencer K. and McGibbon, Jeremy and Dresdner, Gideon and Kashinath, Karthik and Bonev, Boris and Pritchard, Michael S. and Bretherton, Christopher S.},
  year = 2024,
  month = sep,
  journal = {Journal of Geophysical Research: Machine Learning and Computation},
  volume = {1},
  number = {3},
  pages = {e2024JH000136},
  issn = {2993-5210},
  doi = {10.1029/2024JH000136},
  urldate = {2024-10-07},
  copyright = {\copyright{} 2024 Allen Institute for Artificial Intelligence. Journal of Geophysical Research: Machine Learning and Computation published by Wiley Periodicals LLC on behalf of American Geophysical Union.},
  langid = {english}
}

@article{Dun2026,
  title = {{{SamudrACE}}: {{Fast}} and {{Accurate Coupled Climate Modeling With 3D Ocean}} and {{Atmosphere Emulators}}},
  shorttitle = {{{SamudrACE}}},
  author = {Duncan, James P. C. and Wu, Elynn and Dheeshjith, Surya and Subel, Adam and Arcomano, Troy and Clark, Spencer K. and Henn, Brian and Kwa, Anna and McGibbon, Jeremy and Perkins, W. Andre and Gregory, William and {Fernandez-Granda}, Carlos and Busecke, Julius and {Watt-Meyer}, Oliver and Hurlin, William J. and Adcroft, Alistair and Zanna, Laure and Bretherton, Christopher},
  year = 2026,
  journal = {Geophysical Research Letters},
  volume = {53},
  number = {11},
  pages = {e2025GL119340},
  issn = {1944-8007},
  doi = {10.1029/2025GL119340},
  urldate = {2026-06-04},
  copyright = {\copyright{} 2026. Allen Institute for Artificial Intelligence. This article has been contributed to by U.S. Government employees and their work is in the public domain in the USA.},
  langid = {english}
}

@article{Eyr2016,
  title = {Overview of the {{Coupled Model Intercomparison Project Phase}} 6 ({{CMIP6}}) experimental design and organization},
  author = {Eyring, Veronika and Bony, Sandrine and Meehl, Gerald A. and Senior, Catherine A. and Stevens, Bjorn and Stouffer, Ronald J. and Taylor, Karl E.},
  year = 2016,
  month = may,
  journal = {Geoscientific Model Development},
  volume = {9},
  number = {5},
  pages = {1937--1958},
  publisher = {Copernicus GmbH},
  issn = {1991-959X},
  doi = {10.5194/gmd-9-1937-2016},
  urldate = {2023-10-02},
  langid = {english}
}

@misc{Gre2026,
  title = {{{FloeNet}}: {{A}} mass-conserving global sea ice emulator that generalizes across climates},
  shorttitle = {{{FloeNet}}},
  author = {Gregory, William and Bushuk, Mitchell and Duncan, James and Wu, Elynn and Subel, Adam and Clark, Spencer K. and Hurlin, Bill and {Watt-Meyer}, Oliver and Adcroft, Alistair and Bretherton, Chris and Zanna, Laure},
  year = 2026,
  month = mar,
  eprint = {2603.12449},
  primaryclass = {physics},
  publisher = {arXiv},
  doi = {10.48550/arXiv.2603.12449},
  urldate = {2026-05-01},
  archiveprefix = {arXiv}
}

@article{Har2020,
  title = {{{GFDL SHiELD}}: {{A Unified System}} for {{Weather-to-Seasonal Prediction}}},
  shorttitle = {{{GFDL SHiELD}}},
  author = {Harris, Lucas and Zhou, Linjiong and Lin, Shian-Jiann and Chen, Jan-Huey and Chen, Xi and Gao, Kun and Morin, Matthew and Rees, Shannon and Sun, Yongqiang and Tong, Mingjing and Xiang, Baoqiang and Bender, Morris and Benson, Rusty and Cheng, Kai-Yuan and Clark, Spencer and Elbert, Oliver D. and Hazelton, Andrew and Huff, J. Jacob and Kaltenbaugh, Alex and Liang, Zhi and Marchok, Timothy and Shin, Hyeyum Hailey and Stern, William},
  year = 2020,
  month = sep,
  journal = {Journal of Advances in Modeling Earth Systems},
  volume = {12},
  number = {10},
  pages = {e2020MS002223},
  issn = {1942-2466},
  doi = {10.1029/2020MS002223},
  urldate = {2020-11-16},
  copyright = {\copyright 2020. The Authors.},
  langid = {english}
}

@article{Har2023,
  title = {A {{Global Survey}} of {{Rotating Convective Updrafts}} in the {{GFDL X-SHiELD}} 2021 {{Global Storm Resolving Model}}},
  author = {Harris, Lucas and Zhou, Linjiong and Kaltenbaugh, Alex and Clark, Spencer and Cheng, Kai-Yuan and Bretherton, Chris},
  year = 2023,
  journal = {Journal of Geophysical Research: Atmospheres},
  volume = {128},
  number = {10},
  pages = {e2022JD037823},
  issn = {2169-8996},
  doi = {10.1029/2022JD037823},
  urldate = {2025-10-20},
  copyright = {\copyright{} 2023. American Geophysical Union. All Rights Reserved.},
  langid = {english}
}

@misc{Hen2026,
  title = {{{AIMIP Phase}} 1: systematic evaluations of {{AI}} weather and climate models},
  shorttitle = {{{AIMIP Phase}} 1},
  author = {Henn, Brian and Bretherton, Christopher S. and Kodunov, Nikolay and Lessig, Christian and Molina, Maria J. and Arcomano, Troy and {Watt-Meyer}, Oliver and Couairon, Guillaume and Singh, Renu and Brunstein, Robert and Hasson, Yana and Jost, Antonia and Brenowitz, Noah and Manshausen, Peter and {Cresswell-Clay}, Nathaniel and Durran, Dale and Hall, Kyle Joseph Chen and Yuval, Janni and Kochkov, Dmitrii and Hoyer, Stephan and {Lopez-Gomez}, Ignacio},
  year = 2026,
  month = may,
  publisher = {arXiv},
  doi = {10.48550/arXiv.2605.06944},
  urldate = {2026-05-11},
  langid = {english}
}

@article{Her2020,
  title = {The {{ERA5}} global reanalysis},
  author = {Hersbach, Hans and Bell, Bill and Berrisford, Paul and Hirahara, Shoji and Hor{\'a}nyi, Andr{\'a}s and {Mu{\~n}oz-Sabater}, Joaqu{\'i}n and Nicolas, Julien and Peubey, Carole and Radu, Raluca and Schepers, Dinand and Simmons, Adrian and Soci, Cornel and Abdalla, Saleh and Abellan, Xavier and Balsamo, Gianpaolo and Bechtold, Peter and Biavati, Gionata and Bidlot, Jean and Bonavita, Massimo and De Chiara, Giovanna and Dahlgren, Per and Dee, Dick and Diamantakis, Michail and Dragani, Rossana and Flemming, Johannes and Forbes, Richard and Fuentes, Manuel and Geer, Alan and Haimberger, Leo and Healy, Sean and Hogan, Robin J. and H{\'o}lm, El{\'i}as and Janiskov{\'a}, Marta and Keeley, Sarah and Laloyaux, Patrick and Lopez, Philippe and Lupu, Cristina and Radnoti, Gabor and {de Rosnay}, Patricia and Rozum, Iryna and Vamborg, Freja and Villaume, Sebastien and Th{\'e}paut, Jean-No{\"e}l},
  year = 2020,
  journal = {Quarterly Journal of the Royal Meteorological Society},
  volume = {146},
  number = {730},
  pages = {1999--2049},
  issn = {1477-870X},
  doi = {10.1002/qj.3803},
  urldate = {2025-10-14},
  copyright = {\copyright{} 2020 The Authors. Quarterly Journal of the Royal Meteorological Society published by John Wiley \& Sons Ltd on behalf of the Royal Meteorological Society.},
  langid = {english}
}

@article{Hua2014a,
  title = {Why logarithmic? {{A}} note on the dependence of radiative forcing on gas concentration},
  shorttitle = {Why logarithmic?},
  author = {Huang, Yi and Bani Shahabadi, Maziar},
  year = 2014,
  month = nov,
  journal = {Journal of Geophysical Research: Atmospheres},
  volume = {119},
  number = {24},
  pages = {13,683--13,689},
  issn = {2169-8996},
  doi = {10.1002/2014JD022466},
  urldate = {2024-11-16},
  langid = {english}
}

@article{Kam2013,
  title = {Tropospheric adjustment to increasing {{CO2}}: its timescale and the role of land--sea contrast},
  shorttitle = {Tropospheric adjustment to increasing {{CO2}}},
  author = {Kamae, Youichi and Watanabe, Masahiro},
  year = 2013,
  month = dec,
  journal = {Climate Dynamics},
  volume = {41},
  number = {11},
  pages = {3007--3024},
  issn = {1432-0894},
  doi = {10.1007/s00382-012-1555-1},
  urldate = {2026-04-15},
  langid = {english}
}

@article{Kam2015,
  title = {Rapid {{Adjustments}} of {{Cloud}} and {{Hydrological Cycle}} to {{Increasing CO2}}: a {{Review}}},
  shorttitle = {Rapid {{Adjustments}} of {{Cloud}} and {{Hydrological Cycle}} to {{Increasing CO2}}},
  author = {Kamae, Youichi and Watanabe, Masahiro and Ogura, Tomoo and Yoshimori, Masakazu and Shiogama, Hideo},
  year = 2015,
  month = jun,
  journal = {Current Climate Change Reports},
  volume = {1},
  number = {2},
  pages = {103--113},
  issn = {2198-6061},
  doi = {10.1007/s40641-015-0007-5},
  urldate = {2025-11-18},
  langid = {english}
}

@article{Kie2006,
  title = {The {{Climate Sensitivity}} of the {{Community Climate System Model Version}} 3 ({{CCSM3}})},
  author = {Kiehl, Jeffrey T. and Shields, Christine A. and Hack, James J. and Collins, William D.},
  year = 2006,
  month = jun,
  journal = {Journal of Climate},
  volume = {19},
  number = {11},
  pages = {2584--2596},
  publisher = {American Meteorological Society},
  issn = {0894-8755, 1520-0442},
  doi = {10.1175/JCLI3747.1},
  urldate = {2025-05-17},
  chapter = {Journal of Climate},
  langid = {english}
}

@article{Koc2024,
  title = {Neural general circulation models for weather and climate},
  author = {Kochkov, Dmitrii and Yuval, Janni and Langmore, Ian and Norgaard, Peter and Smith, Jamie and Mooers, Griffin and Kl{\"o}wer, Milan and Lottes, James and Rasp, Stephan and D{\"u}ben, Peter and Hatfield, Sam and Battaglia, Peter and {Sanchez-Gonzalez}, Alvaro and Willson, Matthew and Brenner, Michael P. and Hoyer, Stephan},
  year = 2024,
  month = aug,
  journal = {Nature},
  volume = {632},
  number = {8027},
  pages = {1060--1066},
  publisher = {Nature Publishing Group},
  issn = {1476-4687},
  doi = {10.1038/s41586-024-07744-y},
  urldate = {2024-09-11},
  copyright = {2024 The Author(s)},
  langid = {english}
}

@article{Lan2026,
  title = {Forecasting the {{Future With Yesterday}}'s {{Climate}}: {{Temperature Bias}} in {{AI Weather}} and {{Climate Models}}},
  shorttitle = {Forecasting the {{Future With Yesterday}}'s {{Climate}}},
  author = {Landsberg, Jacob B. and Barnes, Elizabeth A.},
  year = 2026,
  journal = {Geophysical Research Letters},
  volume = {53},
  number = {6},
  pages = {e2025GL119740},
  issn = {1944-8007},
  doi = {10.1029/2025GL119740},
  urldate = {2026-04-07},
  copyright = {\copyright{} 2026. The Author(s).},
  langid = {english}
}

@article{Lan2026b,
  title = {{{AIFS-CRPS}}: ensemble forecasting using a model trained with a loss function based on the continuous ranked probability score},
  shorttitle = {{{AIFS-CRPS}}},
  author = {Lang, Simon and Alexe, Mihai and Clare, Mariana C. A. and Roberts, Christopher and Adewoyin, Rilwan and Ben Bouall{\`e}gue, Zied and Chantry, Matthew and Dramsch, Jesper and Dueben, Peter D. and Hahner, Sara and Maciel, Pedro and {Prieto-Nemesio}, Ana and O'Brien, Cathal and Pinault, Florian and Polster, Jan and Raoult, Baudouin and Tietsche, Steffen and Leutbecher, Martin},
  year = 2026,
  month = feb,
  journal = {npj Artificial Intelligence},
  volume = {2},
  number = {1},
  pages = {18},
  publisher = {Nature Publishing Group},
  issn = {3005-1460},
  doi = {10.1038/s44387-026-00073-7},
  urldate = {2026-06-05},
  copyright = {2026 The Author(s)},
  langid = {english}
}

@misc{Lev2026,
  title = {On the seasonal predictability of the 2020 {{North Atlantic}} tropical cyclone season},
  author = {Levin, Emma Lilly and Chien, Mu-Ting and Barnes, Elizabeth and He, Haozhe and Vecchi, Gabriel and Yang, Wenchang},
  year = 2026,
  month = mar,
  publisher = {EarthArXiv},
  doi = {10.31223/X5CN1R},
  urldate = {2026-03-27},
  copyright = {No Creative Commons license},
  langid = {english}
}

@misc{Mah2026,
  title = {Examining {{Fast Radiative Feedbacks Using Machine-Learning Weather Emulators}}},
  author = {Mahesh, Ankur and Collins, William D. and O'Brien, Travis A. and Goddard, Paul B. and Zebaze, Sinclaire and Subramanian, Shashank and Duncan, James P. C. and {Watt-Meyer}, Oliver and Bonev, Boris and Kurth, Thorsten and Kashinath, Karthik and Pritchard, Michael S. and Yang, Da},
  year = 2026,
  month = feb,
  eprint = {2602.16090},
  primaryclass = {physics},
  publisher = {arXiv},
  doi = {10.48550/arXiv.2602.16090},
  urldate = {2026-03-27},
  archiveprefix = {arXiv}
}

@misc{McG2026,
  title = {ai2cm/ace: 2026.5.1},
  shorttitle = {ai2cm/ace},
  author = {McGibbon, Jeremy and {Watt-Meyer}, Oliver and Duncan, James and Kwa, Anna and Henn, Brian and Clark, Spencer and Wu, Elynn and Perkins, W. Andre and Arcomano, Troy and {rebassoo} and Dodson, Anna and Mahfouz, Naser and Dheeshjith, Surya and Gregory, Will and Yik, William and Dresdner, Gideon and Pathak, Jaideep},
  year = 2026,
  month = may,
  doi = {10.5281/zenodo.20128254},
  urldate = {2026-06-02},
  howpublished = {Zenodo}
}

@article{Mei2020,
  title = {The shared socio-economic pathway ({{SSP}}) greenhouse gas concentrations and their extensions to 2500},
  author = {Meinshausen, Malte and Nicholls, Zebedee R. J. and Lewis, Jared and Gidden, Matthew J. and Vogel, Elisabeth and Freund, Mandy and Beyerle, Urs and Gessner, Claudia and Nauels, Alexander and Bauer, Nico and Canadell, Josep G. and Daniel, John S. and John, Andrew and Krummel, Paul B. and Luderer, Gunnar and Meinshausen, Nicolai and Montzka, Stephen A. and Rayner, Peter J. and Reimann, Stefan and Smith, Steven J. and {van den Berg}, Marten and Velders, Guus J. M. and Vollmer, Martin K. and Wang, Ray H. J.},
  year = 2020,
  month = aug,
  journal = {Geoscientific Model Development},
  volume = {13},
  number = {8},
  pages = {3571--3605},
  publisher = {Copernicus GmbH},
  issn = {1991-959X},
  doi = {10.5194/gmd-13-3571-2020},
  urldate = {2026-05-01},
  langid = {english}
}

@article{Mer2024,
  title = {Climate sensitivity and relative humidity changes in global storm-resolving model simulations of climate change},
  author = {Merlis, Timothy M. and Cheng, Kai-Yuan and Guendelman, Ilai and Harris, Lucas and Bretherton, Christopher S. and Bolot, Maximilien and Zhou, Linjiong and Kaltenbaugh, Alex and Clark, Spencer K. and Vecchi, Gabriel A. and Fueglistaler, Stephan},
  year = 2024,
  month = jun,
  journal = {Science Advances},
  volume = {10},
  number = {26},
  pages = {eadn5217},
  publisher = {American Association for the Advancement of Science},
  doi = {10.1126/sciadv.adn5217},
  urldate = {2026-03-31}
}

@article{Nic2020,
  title = {Reduced {{Complexity Model Intercomparison Project Phase}} 1: introduction and evaluation of global-mean temperature response},
  shorttitle = {Reduced {{Complexity Model Intercomparison Project Phase}} 1},
  author = {Nicholls, Zebedee R. J. and Meinshausen, Malte and Lewis, Jared and Gieseke, Robert and Dommenget, Dietmar and Dorheim, Kalyn and Fan, Chen-Shuo and Fuglestvedt, Jan S. and Gasser, Thomas and Gol{\"u}ke, Ulrich and Goodwin, Philip and Hartin, Corinne and Hope, Austin P. and Kriegler, Elmar and Leach, Nicholas J. and Marchegiani, Davide and McBride, Laura A. and Quilcaille, Yann and Rogelj, Joeri and Salawitch, Ross J. and Samset, Bj{\o}rn H. and Sandstad, Marit and Shiklomanov, Alexey N. and Skeie, Ragnhild B. and Smith, Christopher J. and Smith, Steve and Tanaka, Katsumasa and Tsutsui, Junichi and Xie, Zhiang},
  year = 2020,
  month = oct,
  journal = {Geoscientific Model Development},
  volume = {13},
  number = {11},
  pages = {5175--5190},
  publisher = {Copernicus GmbH},
  issn = {1991-959X},
  doi = {10.5194/gmd-13-5175-2020},
  urldate = {2026-06-03},
  langid = {english}
}

@misc{Per2026,
  title = {{{HiRO-ACE}}: {{Fast}} and skillful {{AI}} emulation and downscaling trained on a 3 km global storm-resolving model},
  shorttitle = {{{HiRO-ACE}}},
  author = {Perkins, W. Andre and Kwa, Anna and McGibbon, Jeremy and Arcomano, Troy and Clark, Spencer K. and {Watt-Meyer}, Oliver and Bretherton, Christopher S. and Harris, Lucas M.},
  year = 2026,
  month = feb,
  eprint = {2512.18224},
  primaryclass = {physics},
  publisher = {arXiv},
  doi = {10.48550/arXiv.2512.18224},
  urldate = {2026-03-27},
  archiveprefix = {arXiv}
}

@article{Pin2020,
  title = {Benchmark {{Calculations}} of {{Radiative Forcing}} by {{Greenhouse Gases}}},
  author = {Pincus, Robert and Buehler, Stefan A. and Brath, Manfred and Crevoisier, Cyril and Jamil, Omar and Franklin Evans, K. and Manners, James and Menzel, Raymond L. and Mlawer, Eli J. and Paynter, David and Pernak, Rick L. and Tellier, Yoann},
  year = 2020,
  month = nov,
  journal = {Journal of Geophysical Research: Atmospheres},
  volume = {125},
  number = {23},
  pages = {e2020JD033483},
  issn = {2169-8996},
  doi = {10.1029/2020JD033483},
  urldate = {2024-11-19},
  copyright = {\copyright 2020. The Authors.},
  langid = {english}
}

@article{Ria2017,
  title = {The {{Shared Socioeconomic Pathways}} and their energy, land use, and greenhouse gas emissions implications: {{An}} overview},
  shorttitle = {The {{Shared Socioeconomic Pathways}} and their energy, land use, and greenhouse gas emissions implications},
  author = {Riahi, Keywan and {van Vuuren}, Detlef P. and Kriegler, Elmar and Edmonds, Jae and O'Neill, Brian C. and Fujimori, Shinichiro and Bauer, Nico and Calvin, Katherine and Dellink, Rob and Fricko, Oliver and Lutz, Wolfgang and Popp, Alexander and Cuaresma, Jesus Crespo and Kc, Samir and Leimbach, Marian and Jiang, Leiwen and Kram, Tom and Rao, Shilpa and Emmerling, Johannes and Ebi, Kristie and Hasegawa, Tomoko and Havlik, Petr and Humpen{\"o}der, Florian and Da Silva, Lara Aleluia and Smith, Steve and Stehfest, Elke and Bosetti, Valentina and Eom, Jiyong and Gernaat, David and Masui, Toshihiko and Rogelj, Joeri and Strefler, Jessica and Drouet, Laurent and Krey, Volker and Luderer, Gunnar and Harmsen, Mathijs and Takahashi, Kiyoshi and Baumstark, Lavinia and Doelman, Jonathan C. and Kainuma, Mikiko and Klimont, Zbigniew and Marangoni, Giacomo and {Lotze-Campen}, Hermann and Obersteiner, Michael and Tabeau, Andrzej and Tavoni, Massimo},
  year = 2017,
  month = jan,
  journal = {Global Environmental Change},
  volume = {42},
  pages = {153--168},
  issn = {0959-3780},
  doi = {10.1016/j.gloenvcha.2016.05.009},
  urldate = {2024-10-07}
}

@article{Sah2014,
  title = {The {{NCEP Climate Forecast System Version}} 2},
  author = {Saha, Suranjana and Moorthi, Shrinivas and Wu, Xingren and Wang, Jiande and Nadiga, Sudhir and Tripp, Patrick and Behringer, David and Hou, Yu-Tai and Chuang, Hui-ya and Iredell, Mark and Ek, Michael and Meng, Jesse and Yang, Rongqian and Mendez, Malaqu{\'i}as Pe{\~n}a and van den Dool, Huug and Zhang, Qin and Wang, Wanqiu and Chen, Mingyue and Becker, Emily},
  year = 2014,
  month = mar,
  journal = {Journal of Climate},
  volume = {27},
  number = {6},
  pages = {2185--2208},
  publisher = {American Meteorological Society},
  issn = {0894-8755, 1520-0442},
  doi = {10.1175/JCLI-D-12-00823.1},
  urldate = {2025-05-17},
  chapter = {Journal of Climate},
  langid = {english}
}

@article{Sha2025,
  title = {Improving {{AI Weather Prediction Models Using Global Mass}} and {{Energy Conservation Schemes}}},
  author = {Sha, Yingkai and Schreck, John S. and Chapman, William and Gagne II, David John},
  year = 2025,
  journal = {Journal of Advances in Modeling Earth Systems},
  volume = {17},
  number = {11},
  pages = {e2025MS005138},
  issn = {1942-2466},
  doi = {10.1029/2025MS005138},
  urldate = {2026-04-04},
  copyright = {\copyright{} 2025 The Author(s). Journal of Advances in Modeling Earth Systems published by Wiley Periodicals LLC on behalf of American Geophysical Union.},
  langid = {english}
}

@article{Teb2025a,
  title = {Emulators of {{Climate Model Output}}},
  author = {Tebaldi, C. and Selin, N. E. and Ferrari, R. and Flierl, G.},
  year = 2025,
  month = oct,
  journal = {Annual Review of Environment and Resources},
  volume = {50},
  number = {Volume 50, 2025},
  pages = {709--737},
  publisher = {Annual Reviews},
  issn = {1543-5938, 1545-2050},
  doi = {10.1146/annurev-environ-012125-085838},
  urldate = {2026-03-27},
  langid = {english}
}

@article{Thi2003,
  title = {A {{New High-Resolution Blended Real-Time Global Sea Surface Temperature Analysis}}},
  author = {Thi{\'e}baux, Jean and Rogers, Eric and Wang, Wanqiu and Katz, Bert},
  year = 2003,
  month = may,
  journal = {Bulletin of the American Meteorological Society},
  volume = {84},
  number = {5},
  pages = {645--656},
  publisher = {American Meteorological Society},
  issn = {0003-0007, 1520-0477},
  doi = {10.1175/BAMS-84-5-645},
  urldate = {2025-05-17},
  chapter = {Bulletin of the American Meteorological Society},
  langid = {english}
}

@misc{Wat2023,
  title = {{{ACE}}: {{A}} fast, skillful learned global atmospheric model for climate prediction},
  shorttitle = {{{ACE}}},
  author = {{Watt-Meyer}, Oliver and Dresdner, Gideon and McGibbon, Jeremy and Clark, Spencer K. and Henn, Brian and Duncan, James and Brenowitz, Noah D. and Kashinath, Karthik and Pritchard, Michael S. and Bonev, Boris and Peters, Matthew E. and Bretherton, Christopher S.},
  year = 2023,
  month = dec,
  eprint = {2310.02074},
  primaryclass = {physics},
  publisher = {arXiv},
  doi = {10.48550/arXiv.2310.02074},
  urldate = {2025-05-15},
  archiveprefix = {arXiv}
}

@article{Wat2025,
  title = {{{ACE2}}: accurately learning subseasonal to decadal atmospheric variability and forced responses},
  shorttitle = {{{ACE2}}},
  author = {{Watt-Meyer}, Oliver and Henn, Brian and McGibbon, Jeremy and Clark, Spencer K. and Kwa, Anna and Perkins, W. Andre and Wu, Elynn and Harris, Lucas and Bretherton, Christopher S.},
  year = 2025,
  month = may,
  journal = {npj Climate and Atmospheric Science},
  volume = {8},
  number = {1},
  pages = {205},
  publisher = {Nature Publishing Group},
  issn = {2397-3722},
  doi = {10.1038/s41612-025-01090-0},
  urldate = {2025-07-08},
  copyright = {2025 The Author(s)},
  langid = {english}
}

@article{Web2017,
  title = {The {{Cloud Feedback Model Intercomparison Project}} ({{CFMIP}}) contribution to {{CMIP6}}},
  author = {Webb, Mark J. and Andrews, Timothy and {Bodas-Salcedo}, Alejandro and Bony, Sandrine and Bretherton, Christopher S. and Chadwick, Robin and Chepfer, H{\'e}l{\`e}ne and Douville, Herv{\'e} and Good, Peter and Kay, Jennifer E. and Klein, Stephen A. and Marchand, Roger and Medeiros, Brian and Siebesma, A. Pier and Skinner, Christopher B. and Stevens, Bjorn and Tselioudis, George and Tsushima, Yoko and Watanabe, Masahiro},
  year = 2017,
  month = jan,
  journal = {Geoscientific Model Development},
  volume = {10},
  number = {1},
  pages = {359--384},
  publisher = {Copernicus GmbH},
  issn = {1991-959X},
  doi = {10.5194/gmd-10-359-2017},
  urldate = {2026-04-19},
  langid = {english}
}

@article{Zel2013,
  title = {Contributions of {{Different Cloud Types}} to {{Feedbacks}} and {{Rapid Adjustments}} in {{CMIP5}}},
  author = {Zelinka, Mark D. and Klein, Stephen A. and Taylor, Karl E. and Andrews, Timothy and Webb, Mark J. and Gregory, Jonathan M. and Forster, Piers M.},
  year = 2013,
  month = jul,
  journal = {Journal of Climate},
  volume = {26},
  number = {14},
  pages = {5007--5027},
  publisher = {American Meteorological Society},
  issn = {0894-8755, 1520-0442},
  doi = {10.1175/JCLI-D-12-00555.1},
  urldate = {2026-06-04},
  chapter = {Journal of Climate},
  langid = {english}
}

@misc{Zha2026,
  title = {The {{Equilibrium Response}} of {{Atmospheric Machine-Learning Models}} to {{Uniform Sea Surface Temperature Warming}}},
  author = {Zhang, Bosong and Merlis, Timothy M.},
  year = 2026,
  month = jan,
  eprint = {2510.02415},
  primaryclass = {physics},
  publisher = {arXiv},
  doi = {10.48550/arXiv.2510.02415},
  urldate = {2026-03-27},
  archiveprefix = {arXiv}
}

@article{Zho2019,
  title = {Toward {{Convective-Scale Prediction}} within the {{Next Generation Global Prediction System}}},
  author = {Zhou, Linjiong and Lin, Shian-Jiann and Chen, Jan-Huey and Harris, Lucas M. and Chen, Xi and Rees, Shannon L.},
  year = 2019,
  month = jul,
  journal = {Bulletin of the American Meteorological Society},
  volume = {100},
  number = {7},
  pages = {1225--1243},
  publisher = {American Meteorological Society},
  issn = {0003-0007, 1520-0477},
  doi = {10.1175/BAMS-D-17-0246.1},
  urldate = {2026-06-04},
  chapter = {Bulletin of the American Meteorological Society},
  langid = {english}
}

@article{Zho2022a,
  title = {Integrated {{Dynamics-Physics Coupling}} for {{Weather}} to {{Climate Models}}: {{GFDL SHiELD With In-Line Microphysics}}},
  shorttitle = {Integrated {{Dynamics-Physics Coupling}} for {{Weather}} to {{Climate Models}}},
  author = {Zhou, Linjiong and Harris, Lucas},
  year = 2022,
  journal = {Geophysical Research Letters},
  volume = {49},
  number = {21},
  pages = {e2022GL100519},
  issn = {1944-8007},
  doi = {10.1029/2022GL100519},
  urldate = {2026-04-02},
  copyright = {\copyright{} 2022. The Authors.},
  langid = {english}
}
%TC:endignore
% don't specify bibliographystyle
%
%%%%%%%%%%%%%%%%%%%%%%%%%%%%%%%%%%%%%%%%%%%%%%%

%\bibliography{ enter your bibtex bibliography filename here }

%Reference citation instructions and examples:
%
% Please use ONLY \cite and \citeA for reference citations.
% \cite for parenthetical references
% ...as shown in recent studies (Simpson et al., 2019)
% \citeA for in-text citations
% ...Simpson et al. (2019) have shown...
%
%
%...as shown by \citeA{jskilby}.
%...as shown by \citeA{lewin76}, \citeA{carson86}, \citeA{bartoldy02}, and \citeA{rinaldi03}.
%...has been shown \cite{jskilbye}.
%...has been shown \cite{lewin76,carson86,bartoldy02,rinaldi03}.
%... \cite <i.e.>[]{lewin76,carson86,bartoldy02,rinaldi03}.
%...has been shown by \cite <e.g.,>[and others]{lewin76}.
%
% apacite uses < > for prenotes and [ ] for postnotes
% DO NOT use other cite commands (e.g., \citet, \citep, \citeyear, \nocite, \citealp, etc.).
%

\end{document}

%% ------------------------------------------------------------------------ %%
%
%  TITLE
%
%% ------------------------------------------------------------------------ %%

%\includegraphics{agu_pubart-white_reduced.eps}

\title{Supporting Information for ``Disentangling the effects of sea surface temperature and CO$_2$ in global machine learned weather-climate emulators''}
%
% e.g., \title{Supporting Information for "Terrestrial ring current:
% Origin, formation, and decay $\alpha\beta\Gamma\Delta$"}
%
%DOI: 10.1002/%insert paper number here%

%% ------------------------------------------------------------------------ %%
%
%  AUTHORS AND AFFILIATIONS
%
%% ------------------------------------------------------------------------ %%

% List authors by first name or initial followed by last name and
% separated by commas. Use \affil{} to number affiliations, and
% \thanks{} for author notes.
% Additional author notes should be indicated with \thanks{} (for
% example, for current addresses).

% Example: \authors{A. B. Author\affil{1}\thanks{Current address, Antartica}, B. C. Author\affil{2,3}, and D. E.
% Author\affil{3,4}\thanks{Also funded by Monsanto.}}

\authors{Spencer K. Clark\affil{1,2}, Troy Arcomano\affil{1}, James P. C. Duncan\affil{1}, Brian Henn\affil{1}, Anna Kwa\affil{1}, Jeremy McGibbon\affil{1}, W. Andre Perkins\affil{1}, Elynn Wu\affil{1}, Lucas M. Harris\affil{2}, Oliver Watt-Meyer\affil{1}, and Christopher S. Bretherton\affil{1}}

% \affiliation{1}{First Affiliation}
% \affiliation{2}{Second Affiliation}
% \affiliation{3}{Third Affiliation}
% \affiliation{4}{Fourth Affiliation}

\affiliation{1}{Allen Institute for Artificial Intelligence, Seattle, WA}
\affiliation{2}{NOAA/Geophysical Fluid Dynamics Laboratory, Princeton, NJ}

%% ------------------------------------------------------------------------ %%
%
%  BEGIN ARTICLE
%
%% ------------------------------------------------------------------------ %%

% The body of the article must start with a \begin{article} command
%
% \end{article} must follow the references section, before the figures
%  and tables.

\begin{article}

%% ------------------------------------------------------------------------ %%
%
%  TEXT
%
%% ------------------------------------------------------------------------ %%

\noindent\textbf{Contents of this file}
%%%Remove or add items as needed%%%
\begin{enumerate}
\item Text S1
\item Figures S1 to S10
% \item Tables S1 to Sx
%if Tables are larger than 1 page, upload as separate excel file
\end{enumerate}

\end{article}
\clearpage
\noindent\textbf{Text S1: Precipitation power spectrum and probability density function in the 3xCO$_2$ equilibrium climate}

\begin{figure}
\noindent\includegraphics[width=\textwidth]{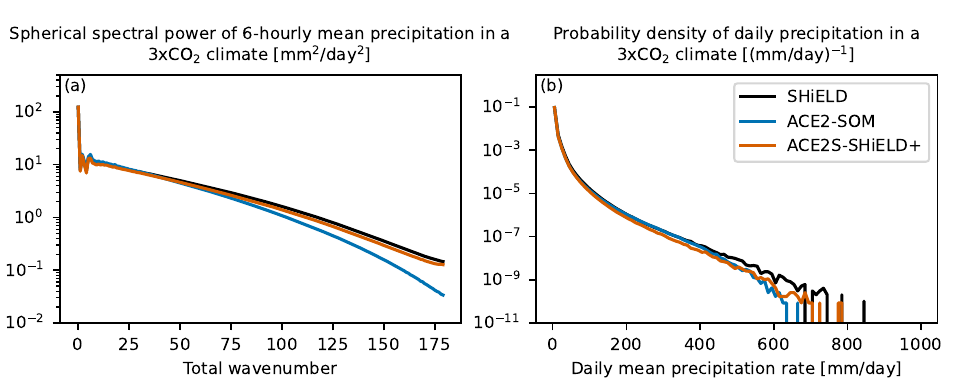}
\caption{Spherical power spectrum of 6-hourly mean precipitation (a) and probability density function of daily mean precipitation in an ensemble of \num{5} \num{10}-year SOM-coupled 3xCO$_2$ equilibrium climate simulations with SHiELD, ACE2-SOM, and ACE2S-SHiELD+ (red).}
\label{fig:3xCO2-PRATEsfc-spectra-and-histograms}
\end{figure}

Figure~\ref{fig:3xCO2-PRATEsfc-spectra-and-histograms}a shows that as a result of the introduction of stochasticity and training with a probabilistic loss function, ACE2S-SHiELD+ produces substantially more power in its predictions of precipitation at the smallest spatial scales than ACE2-SOM, which has a smooth bias due its mean square error loss.  This increased power at the smallest scales results in a slightly improved representation of the most extreme daily-mean precipitation events, i.e. those with rates over \SI{600}{\mm \per \day}, but also too high a frequency of extremely light precipitation events at the expense of more moderate events in the middle of the distribution, as shown in Figure~\ref{fig:3xCO2-PRATEsfc-spectra-and-histograms}b.
\clearpage

\begin{figure}
\noindent\includegraphics[width=\textwidth]{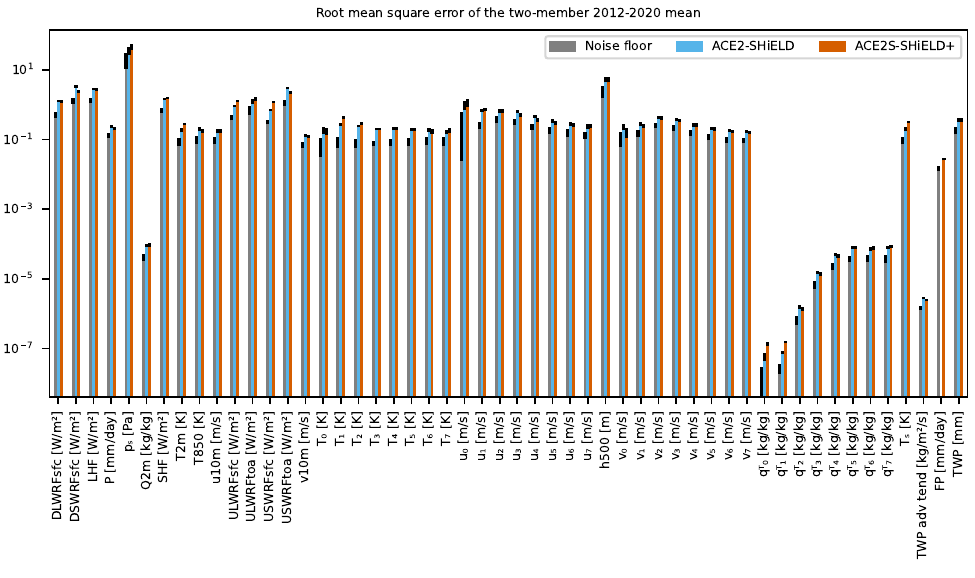}
\caption{Global root mean square error of the 2012 through 2020 mean in AMIP inference for each variable predicted by ACE2-SHiELD and ACE2S-SHiELD+.  Error is computed based on the mean of two ensemble members each with ACE and SHiELD.  The gray bars and black error bars represent the noise floor and its uncertainty, which are computed following a similar approach to \citeA{Cla2025b}.  The noise floor provides an estimate for the expected error of two more SHiELD ensemble members relative to the existing two and can be thought of as a lower bound on emulation error due to internal variability.}
\label{fig:comprehensive-amip-skill}
\end{figure}

\begin{figure}
\noindent\includegraphics[width=\textwidth]{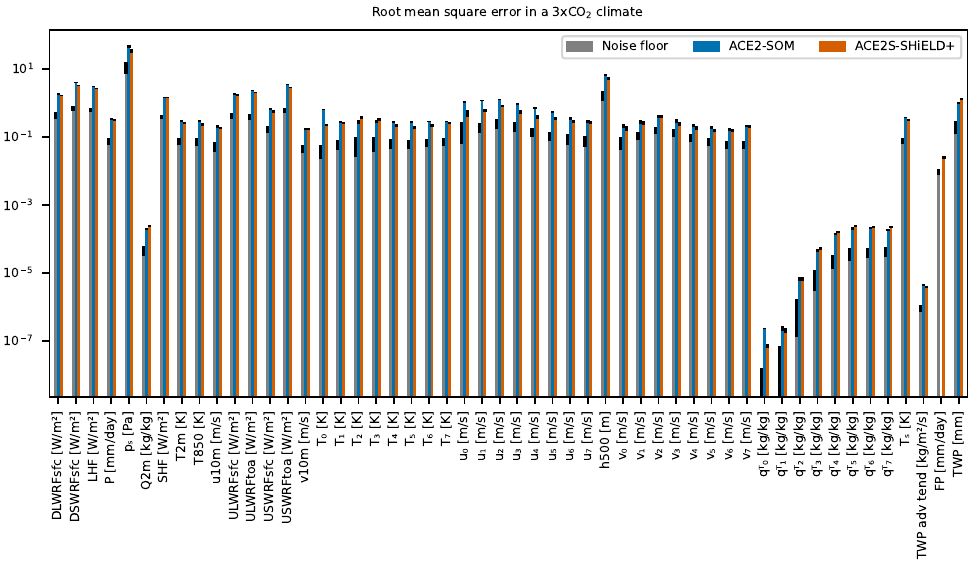}
\caption{Global root mean square error ensemble and time mean in SOM-coupled 3xCO$_2$ equilibrium-climate inference for each variable predicted by ACE2-SOM and ACE2S-SHiELD+.  Error is computed based on the mean of five \num{10}-year ensemble members each with ACE and SHiELD.  The gray bars and black error bars represent the noise floor and its uncertainty, which are computed following a similar approach to \citeA{Cla2025b}.  The noise floor provides an estimate for the expected error of five more SHiELD ensemble members relative to the existing five and can be thought of as a lower bound on emulation error due to internal variability.}
\label{fig:comprehensive-3xCO2-skill}
\end{figure}

\begin{figure}
\noindent\includegraphics[width=\textwidth]{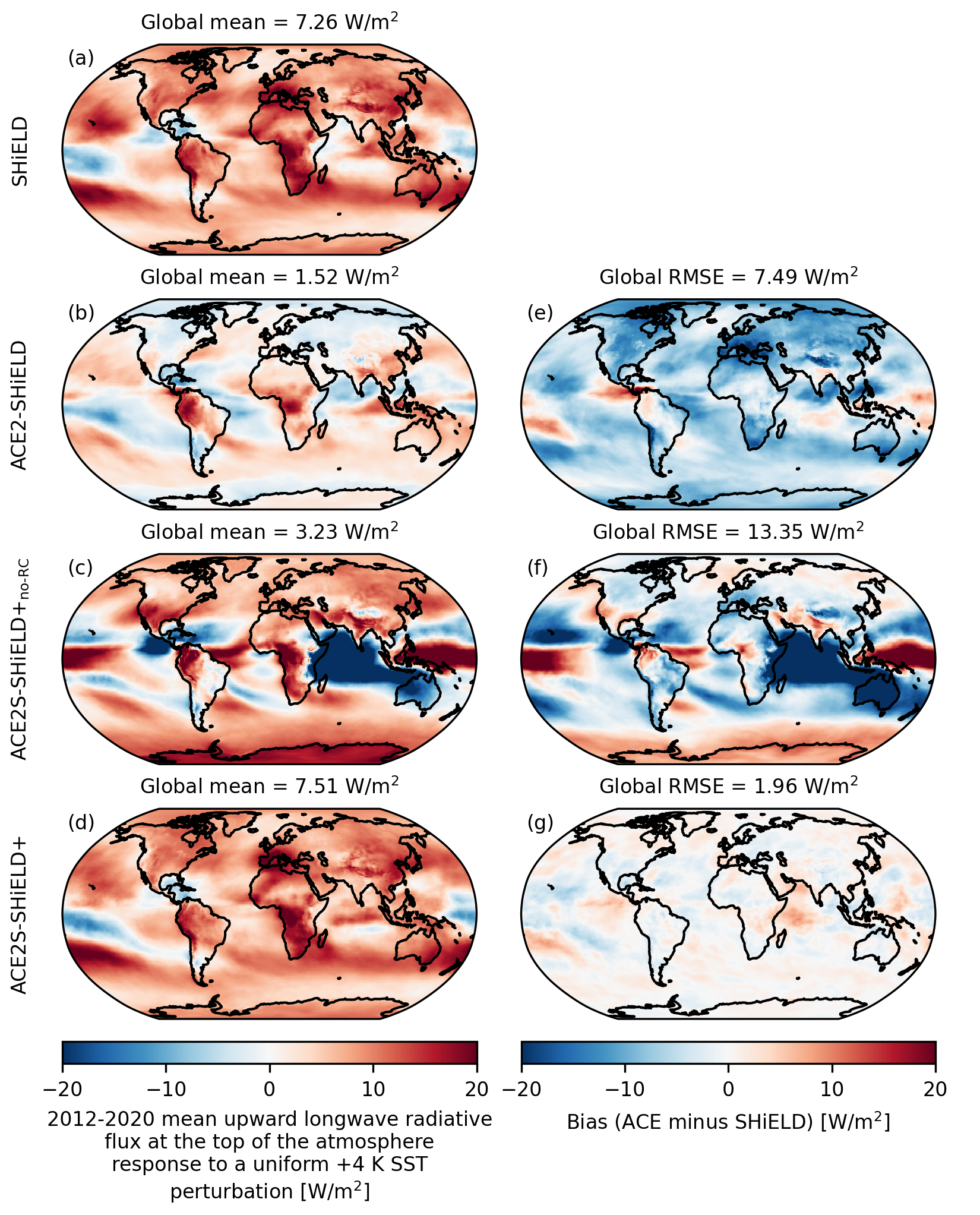}
\caption{As in Figure~6 but for outgoing longwave radiation.}
\label{fig:amip-p4K-ULWRFtoa}
\end{figure}

\begin{figure}
\noindent\includegraphics[width=\textwidth]{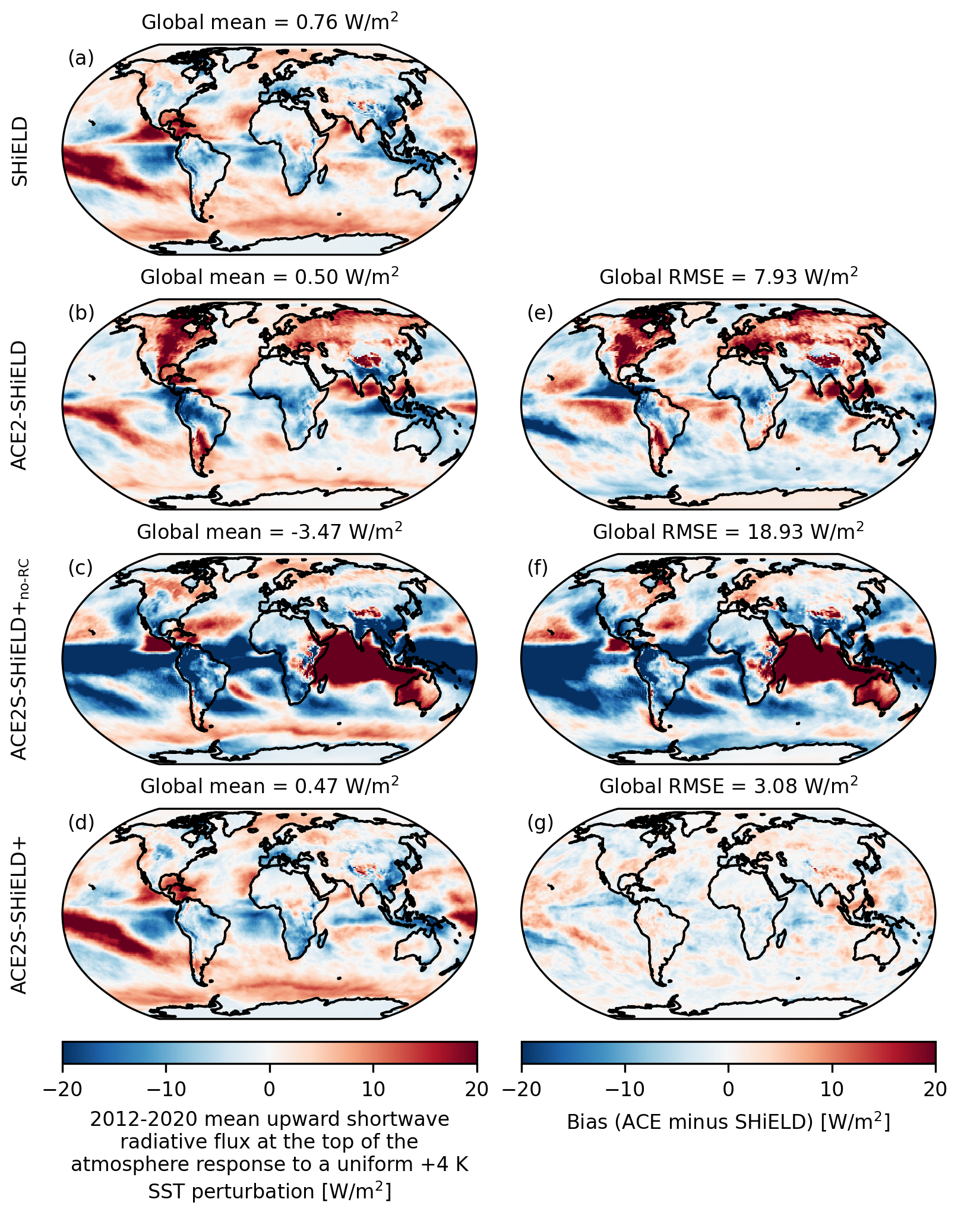}
\caption{As in Figure~6 but for upward shortwave radiative flux at the top of the atmosphere.}
\label{fig:amip-p4K-USWRFtoa}
\end{figure}

\begin{figure}
\noindent\includegraphics[width=\textwidth]{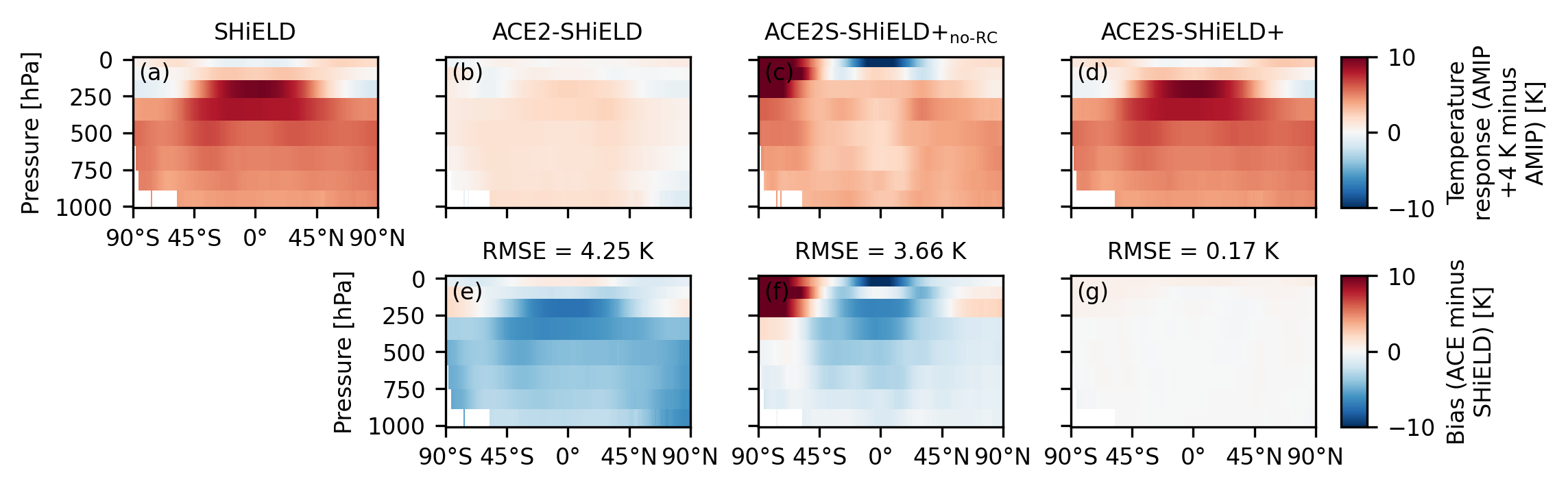}
\caption{2012 through 2020 zonal and time mean difference in air temperature interpolated to surfaces of constant pressure between an AMIP \SI[retain-explicit-plus]{+4}{\K} simulation and an AMIP simulation with SHiELD (a) ACE2S-SHiELD+$_{\text{no-RC}}$ (c), and ACE2S-SHiELD+ (d). Panels (e)-(g) show the response pattern error relative to the target SHiELD; the mass-weighted RMSE is shown in the panel titles.}
\label{fig:amip-p4k-vertical-air-temperature}
\end{figure}

\begin{figure}
\noindent\includegraphics[width=\textwidth]{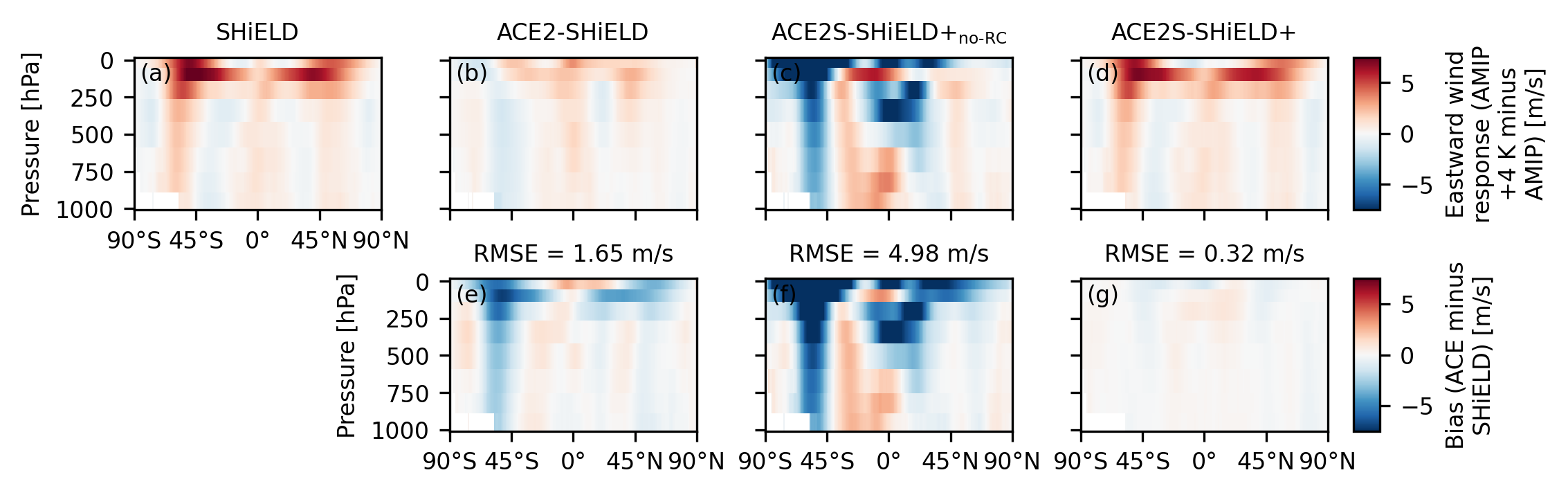}
\caption{As in Figure~\ref{fig:amip-p4k-vertical-air-temperature} but for eastward wind.}
\label{fig:amip-p4k-vertical-eastward-wind}
\end{figure}

\begin{figure}
\noindent\includegraphics[width=\textwidth]{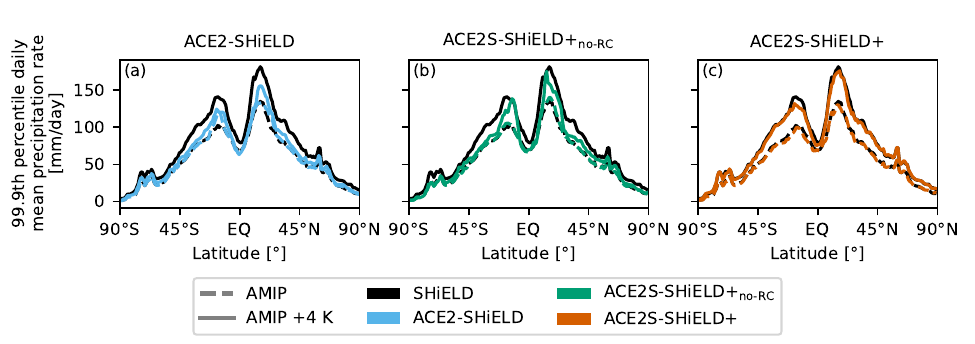}
\caption{99.9th percentile of daily mean precipitation rate from 2012 through 2020 in AMIP (dashed lines) and AMIP \SI[retain-explicit-plus]{+4}{\K} (solid lines) simulations with SHiELD (black), ACE2-SHiELD, ACE2S-SHiELD+$_{\text{no-RC}}$, and ACE2S-SHiELD+.}
\label{fig:amip-p4k-extreme-precipitation}
\end{figure}

\begin{figure}
\noindent\includegraphics[width=\textwidth]{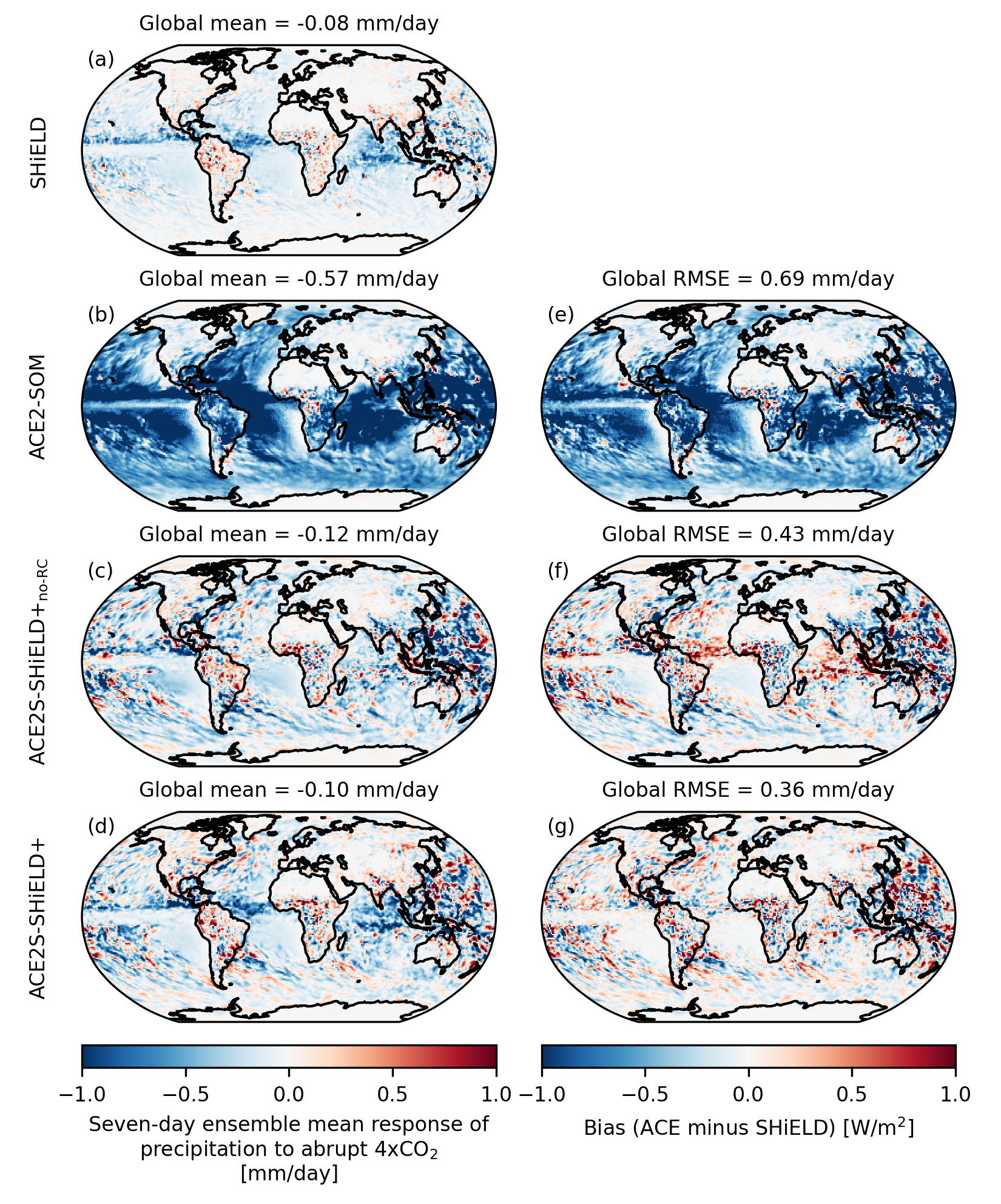}
\caption{As in Figure~9 but for precipitation rate.}
\label{fig:abrupt-4xCO2-PRATEsfc}
\end{figure}

\begin{figure}
\noindent\includegraphics[width=\textwidth]{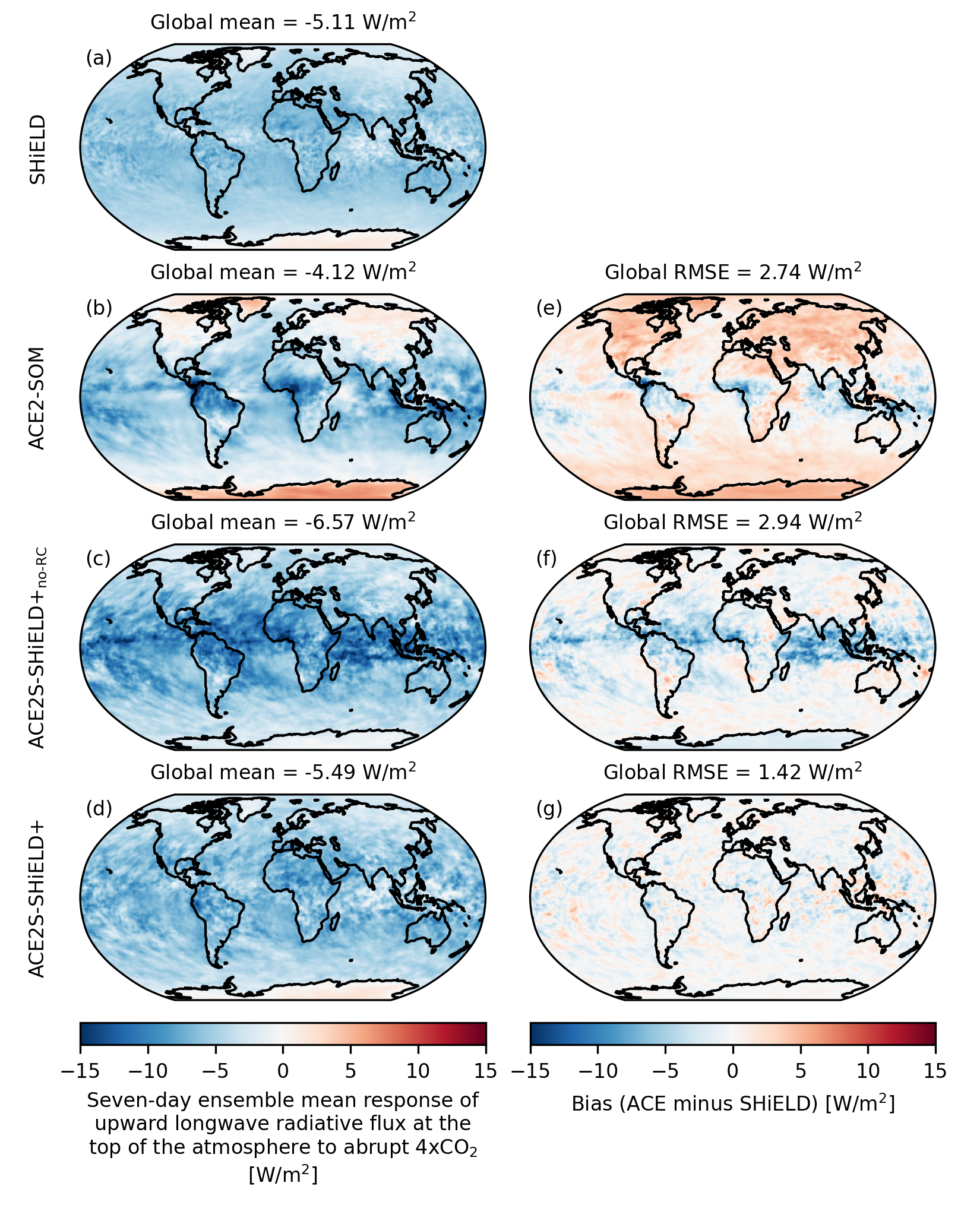}
\caption{As in Figure~9 but for outgoing longwave radiation.}
\label{fig:abrupt-4xCO2-ULWRFtoa}
\end{figure}
\clearpage
\bibliography{references.bib}